\documentclass[10pt,journal,compsoc]{IEEEtran}
\ifCLASSOPTIONcompsoc
  \usepackage[nocompress]{cite}
\else
  \usepackage{cite}
\fi

\usepackage{amsmath,amssymb,amsfonts}
\usepackage[pdftex]{graphicx}
\usepackage[table,xcdraw]{xcolor}
\usepackage{stfloats}
\usepackage{url}
\usepackage{subfigure,stfloats}
\usepackage{booktabs}
\usepackage{hyperref}
\hypersetup{
    colorlinks=true,
    linkcolor=blue,
    filecolor=magenta,      
    urlcolor=cyan,
}

\graphicspath{{Figures/}}

\begin{document}
\title{Pain level and pain-related behaviour classification using GRU-based sparsely-connected RNNs}

\author{\IEEEauthorblockN{Mohammad Mahdi~Dehshibi\IEEEauthorrefmark{1}\IEEEauthorrefmark{4},
Temitayo~Olugbade\IEEEauthorrefmark{2},
Fernando~Diaz-de-Maria\IEEEauthorrefmark{3},
Nadia~Bianchi-Berthouze\IEEEauthorrefmark{2},
Ana~Tajadura-Jim\'{e}nez\IEEEauthorrefmark{1}\IEEEauthorrefmark{2}}\protect\\
\IEEEauthorblockA{\IEEEauthorrefmark{1} Department of Computer Science and Engineering, Universidad Carlos III de Madrid, Legan\'{e}s, Spain.\protect\\}
\IEEEauthorblockA{\IEEEauthorrefmark{2} UCL Interaction Centre, University College London, London, U.K.\protect\\}
\IEEEauthorblockA{\IEEEauthorrefmark{3} Department of Signal Theory and Communications, Universidad Carlos III de Madrid, Legan\'{e}s, Spain.\protect\\}
\IEEEauthorblockA{\IEEEauthorrefmark{4} Unconventional Computing Laboratory, University of the West of England, Bristol, U.K. \protect\\}

\thanks{Corresponding author: M.M.~Dehshibi (mohammad.dehshibi@yahoo.com)}}

% The paper headers
% \markboth{IEEE Journal of Selected Topics in Signal Processing}%
% {Dehshibi \MakeLowercase{\textit{et al.}}: Pain level and pain-related behaviour classification using GRU-based sparsely connected RNNs}

\IEEEtitleabstractindextext{%
\begin{abstract}
There is a growing body of studies on applying deep learning to biometrics analysis. Certain circumstances, however, could impair the objective measures and accuracy of the proposed biometric data analysis methods. For instance, people with chronic pain (CP) unconsciously adapt specific body movements to protect themselves from injury or additional pain. Because there is no dedicated benchmark database to analyse this correlation, we considered one of the specific circumstances that potentially influence a person's biometrics during daily activities in this study and classified pain level and pain-related behaviour in the EmoPain database. To achieve this, we proposed a sparsely-connected recurrent neural networks (s-RNNs) ensemble with the gated recurrent unit (GRU) that incorporates multiple autoencoders using a shared training framework. This architecture is fed by multidimensional data collected from inertial measurement unit (IMU) and surface electromyography (sEMG) sensors. Furthermore, to compensate for variations in the temporal dimension that may not be perfectly represented in the latent space of s-RNNs, we fused hand-crafted features derived from information-theoretic approaches with represented features in the shared hidden state. We conducted several experiments which indicate that the proposed method outperforms the state-of-the-art approaches in classifying both pain level and pain-related behaviour.
\end{abstract}

% Note that keywords are not normally used for peerreview papers.
}

\maketitle
\IEEEraisesectionheading{\section{Introduction}\label{sec:introduction}}
\IEEEPARstart{B}{i}ometric data analysis, which is unique to each individual, has become increasingly popular in recent years for various applications ranging from security~\cite{sundararajan2018deep} to medical diagnosis~\cite{kumar2019deep}. Biometrics analysis is the process of measuring and analysing an individual's physical and behavioural characteristics, such as fingerprints, iris, voice, and body movements, to identify or verify their identity~\cite{jain201650}. 

\begin{figure}[!htbp]
    \centering
    \includegraphics[width=1\linewidth]{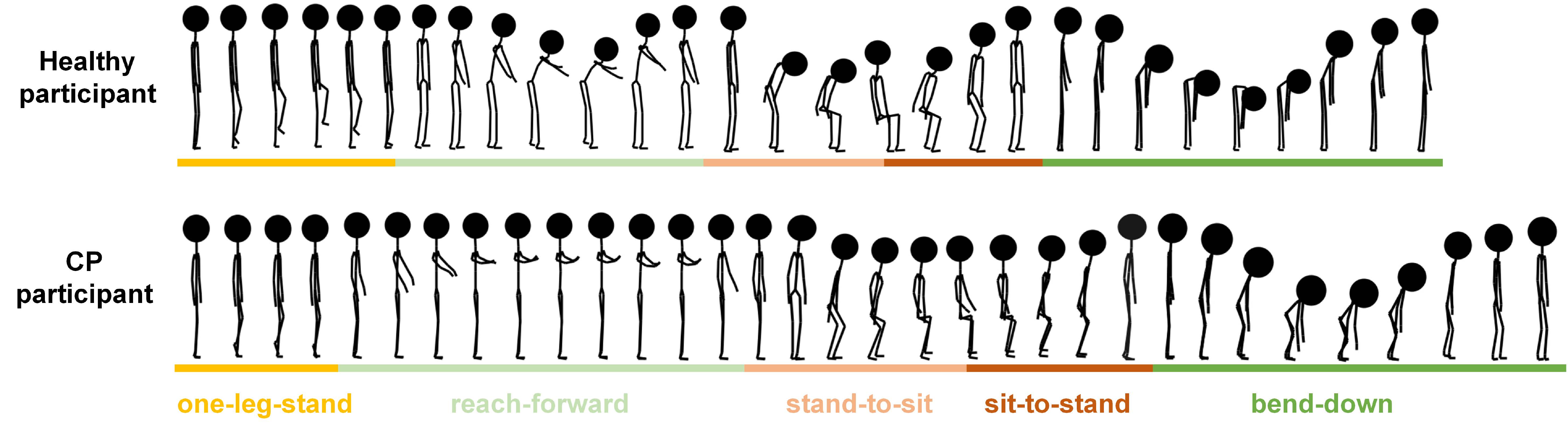}
    \caption{Examples of avatars from 3D joint coordinates data from healthy and CP individuals performing five activities. Image is taken from~\cite{wang2021leveraging}.}
    \label{fig:avatar}
\end{figure}

Deep learning is a type of machine learning that can be used to detect patterns in data automatically. The efficiency of deep learning methods have been shown in various tasks, including image analysis~\cite{dehshibi2019cubic,dehshibi2021deep,adhane2021deep,adhane2021use,ashtari2022amulti}, natural language processing~\cite{otter2020survey,guo2020gluoncv}, and biometrics analysis~\cite{bhanu2017deep}. However, certain circumstances could impair the objective measures and accuracy of the proposed biometric data analysis methods. For example, people suffering from chronic pain (CP) unconsciously adapt body movements to protect themselves from further pain or injury~\cite{keefe1982development,sullivan2006influence}. As a result, they may avoid activities that require bending or lifting. They may also begin sitting for extended periods with a straight back or twisting their trunk instead of bending their body during the sit-to-stand movement to reduce strain on the back muscles. Indeed, the presence of pain and the adoption of protective behaviours can bias the data used to train biometrics analysis algorithms toward non-general bodily movements, reducing the generalizability of these algorithms. For instance, Fig.~\ref{fig:avatar} shows an individual with CP avoids certain activities or strongly alters the way a movement is generally performed, which results in missing data points for those activities.

%---PLACEMENT

As there is no dedicated benchmark database to study the impact of pain presence and protective behaviours in individuals with CP on the performance of biometric analysis algorithms, in this study, we considered one of the particular circumstances that could influence a person's biometrics throughout daily activities. Therefore, we primarily focused on classifying the pain level and pain-related behaviour (\textit{i.e.,} non-protective, protective) in the fully-annotated EmoPain database~\cite{aung2015automatic} and left its application in biometrics for future research.

%---PLACEMENT
\begin{figure*}[!b]
    \centering
    \includegraphics[width=0.95\linewidth]{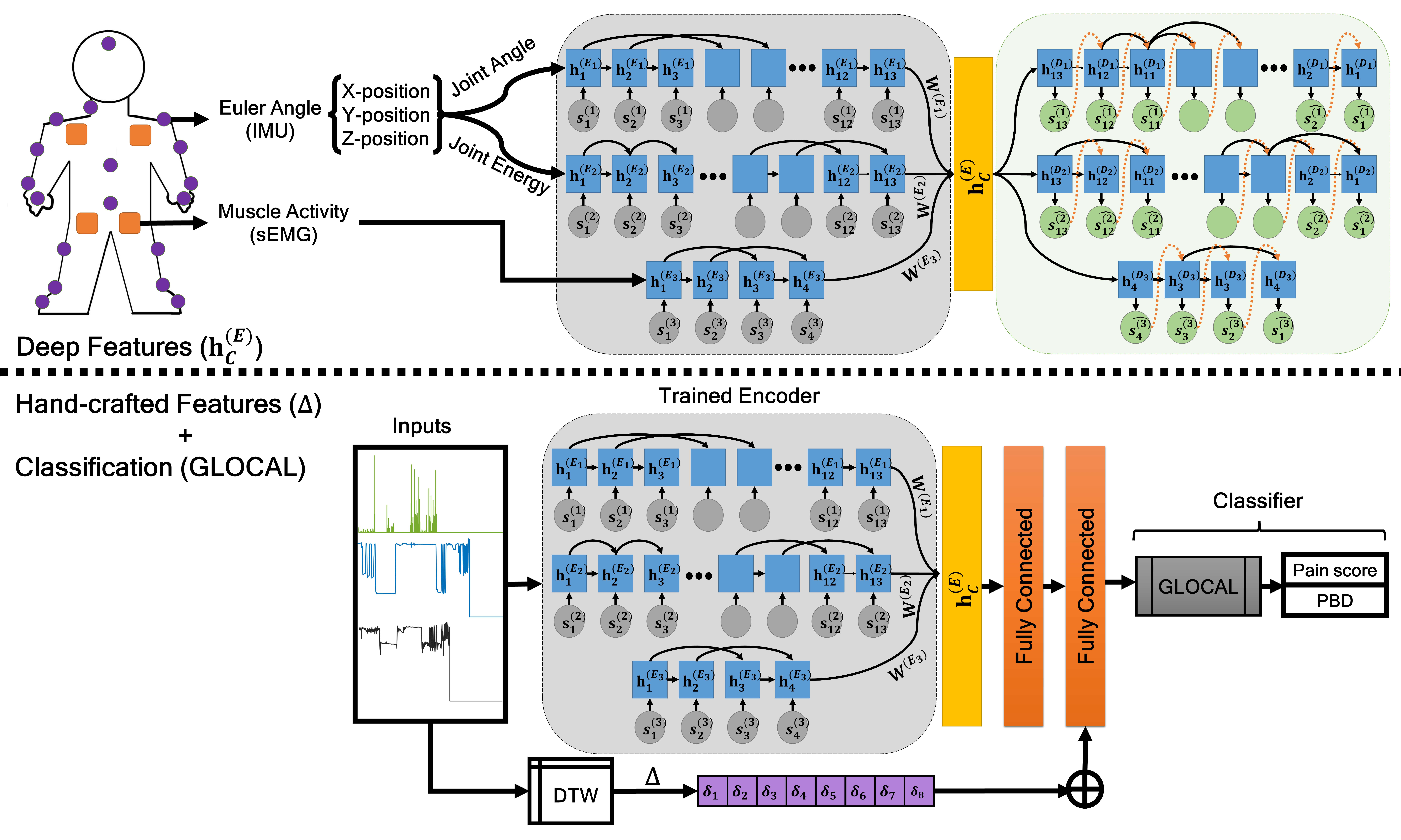}
    \caption{The proposed method pipeline. The IMU and sEMG sensors are placed on the body to record data from participants while they engage in activities. The IMU sensors (violet circle) record $(x, y, z)$ coordinates in the Euler angle for each participant, while the sEMG sensors (orange rectangle) record muscle activity. The positioning of the sensors is presented as an example here that could change depending on the use case explored. The one used in this case study is described in Section~\ref{sec:experiment}. In this schematic, $\mathbf{h}_{C}^{(E)}$ represents the latent space, and $\Delta$ represents extracted hand-crafted features using the information-theoretic approach. In our architecture, hand-crafted features are combined with deep features before being fed into the multi-label and multi-class GLOCAL classifier. As shown in the figure, the labels in our use case are pain level and pain-related behaviour.}
    \label{fig:schematic}
\end{figure*}

We hence argue that the results of this study and the findings of previous studies in this area can pave the way for further research into the effect of protective behaviours on the performance of biometric analysis algorithms and proposing new objective measures for the following reasons: (1) these findings could be used, for instance, to identify areas of a city that are less frequented by people suffering from chronic pain (according to expected statistics) and increase accessibility in such regions; (2) these findings could be used to adapt the computation of people's own biometrics in the context of transitionally altered behaviour to prevalent and fluctuating conditions such as chronic musculoskeletal pain; and (3) identifying these behavioural patterns may assist healthcare practitioners and physiotherapists in designing interventions and providing a physical therapy programme that includes exercises to improve flexibility and strength for people with CP~\cite{hayden2005exercise,koes2006diagnosis}.

The EmoPain database contains data collected from healthy individuals and individuals with CP using wearable inertial measurement unit (IMU) and surface electromyography (sEMG) sensors during exercise movements that reflect activities in everyday life, \textit{e.g.,} sit-to-stand. It is worth noting that the data in the EmoPain database has been labelled with two sets of interest labels: pain level and pain-related behaviour. The characteristics of the EmoPain database pose the following challenges in classification tasks, which we address in the proposed architecture.

\begin{enumerate}
    \item The duration of recorded activities varies per individual, as each individual adjusts their movement patterns based on their needs to minimise energy expenditure while maximising protective behaviour~\cite{collins2015reducing,wang2019learning,wang2021chronic}. The variety in the duration of recorded activities is considerably larger than the number of images in benchmark video databases for human activity recognition (HAR)~\cite{liu2020ntu}. Therefore, modifying deep architectures for HAR from images/videos to classify signals with comparable computational complexity is not straightforward.
    \item Due to the small sample size and class imbalances in the EmoPain database, it is not straightforward, for instance, to combine graph convolutional neural network (GCN) with Long Short-Term Memory (LSTM) networks to simultaneously classify pain level and pain-related behaviour~\cite{wang2021leveraging}.
\end{enumerate}

To address these challenges, we proposed an ensemble of sparsely-connected recurrent neural networks (s-RNNs) with the gated recurrent unit (GRU) that incorporates multiple autoencoders (AEs) using a shared training framework. Multidimensional data feed this architecture to derive representative features from time series. In addition, we benefited from using information-theoretic features~\cite{dehshibi2021electrical,zenil2019causal,wallace1999minimum} to compensate for variations in the temporal dimension that may not be adequately represented in the latent space of s-RNNs. Finally, we used the GLOCAL classifier~\cite{zhu2018multi} to perform the multi-class and multi-label classification (see Fig.~\ref{fig:schematic}). The main contributions of this study can be summarised as follows:

\begin{enumerate}
    \item We proposed an s-RNNs ensemble with GRU in which multiple AEs are incorporated using a shared training framework. In addition, we introduced sparsity into the proposed architecture using recurrent skip connections and auxiliary connections between RNN units. Auxiliary connections help preserve additional hidden states in the past by removing the connection to the immediate previous hidden states. Training with a shared training framework and introducing sparsity to the ensemble of AEs demonstrate their benefit in reducing the probability of overfitting for small-scaled databases and making decoders less vulnerable to outliers.
    \item In order to compensate for variations in the temporal dimension that may not be perfectly represented in the latent space of S-RNNs, we proposed fusing hand-crafted features with those represented features in the shared hidden state before feeding it into the multi-label and multi-class classifier. Our experimental results highlight the cause and benefit of employing information-theoretic features for analysing problems with multidimensional inputs like the one in this study.
    \item We used the GLOCAL classifier to classify pain level and pain-related behaviour jointly, whereas previous studies used different architectures to independently present the results for each task.
    \item We have investigated the use of this novel approach in the case of pain and protective behaviour detection with an in-depth study using the EmoPain database. We conducted comprehensive experiments using several performance metrics to compare the proposed method to the state-of-the-art approaches. Furthermore, we conducted an ablation study to show the effect of each component in the proposed architecture.
\end{enumerate}

Our experimental results demonstrate that the proposed method outperforms the state-of-the-art in classifying both pain level and pain-related behaviour. The rest of this paper is structured as follows: Section~\ref{sec:relatedwork} reviews the previous research. Section~\ref{sec:methodology} provides details on the proposed method. Experimental results and discussions are presented in Section~\ref{sec:experiment}. Finally, the paper is concluded in Section~\ref{sec:conclusion}.
%-------------------------------------------------------------------------

\section{Literature Review}\label{sec:relatedwork}
The use of deep learning to analyse human movement for assessing pain levels and pain-related protective behaviours is a relatively new area of research, and wearable sensors provide a convenient way to gather data for these algorithms. Because studies that specifically look at the impact of protective behaviour on the performance of biometrics analysis have not been conducted due to a lack of a dedicated database for this purpose, we summarise the current state of research analysing human movement data for pain level and pain-related protective behaviour assessment.

Yang et al.~\cite{yang2012machine} proposed a machine learning approach to evaluate the physical performance of individuals with Complex Regional Pain Syndrome (CRPS) using gait data acquired by an accelerometer over short walking distances. The feature set comprises temporal features, gait energy distribution, regularity, and symmetry. They employed a multilayer perceptron neural network, support vector machine, random forest, linear discriminant analysis, and KStar to measure lower back trunk acceleration. They demonstrated that individuals with CRPS alter their gait to protect themselves from further pain. Yoo et al.~\cite{yoo2013interpretation} investigated the relationship between stair ascent movement alterations and pain, radiographic severity, and prognosis of knee osteoarthritis in older women with persistent knee pain. In this research, kinematic predictors of pain were identified using support vector machines (SVM). SVM predictors were stair ascent time, maximal anterior pelvis tilting, knee flexion at initial foot contact, and ankle dorsiflexion at initial foot contact. The result revealed that using machine learning techniques to predict pain and radiographic severity could help researchers better understand risk factors for knee osteoarthritis.

Researchers working on the EmoPain database~\cite{aung2015automatic} found that conventional and deep learning methods for analysing human movement are viable tools for detecting pain-related protective behaviours~\cite{olugbade2015pain,wang2019recurrent,wang2021chronic}. Wang et al. ~\cite{wang2019recurrent,wang2019learning,wang2021chronic} demonstrated that using LSTM-based architectures allows for activity-independent pain-related behaviour detection (PBD) with improved performance. In~\cite{wang2019recurrent,wang2021chronic}, stacked LSTM and dual-stream LSTM were studied for processing of body movement data in conjunction with data augmentation and segmentation window width approaches. Three LSTM layers and two sets of three LSTM layers (for the MoCap and sEMG streams) were used for stacked-LSTM and dual-stream LSTM, respectively. The experimental results revealed that the stacked-LSTM outperforms the dual-stream LSTM and CNN-based models. BodyAttentionNet (BANet), an end-to-end deep learning architecture, was proposed in~\cite{wang2019learning} to perform spontaneous temporal and bodily part subset selection on MoCap data. Attention mechanisms were integrated into the LSTM-based network in the proposed architecture to (1) allow the architecture to focus on the relevant configuration of protective behaviour and (2) reveal how MoCap data without sEMG could help better understand protective behaviour from real-life measurements rather than only lab-based observations. They showed that BANet could achieve favourable performance with less trainable parameters. Although these models were activity-independent and functional across a wide range of activity types, continuous detection was limited to pre-segmented activities of interest, and the association between the kind of activity and protective behaviour was not leveraged.

Olugbade et al.~\cite{olugbade2020movement} proposed a Movement in Multiple Time (MiMT) neural network with distributed time encoding of low-level movement features and joint prediction of pain behaviour across multiple timescales to address these limitations. MiMT could focus on body movement data with independent but coordinating degrees of freedom because it could compute time encoding individually for different sets of anatomical segments using a shared encoder and provide multiple outputs at different timescales for the same label. Wang et al.~\cite{wang2021leveraging} proposed integrating Human Activity Recognition (HAR) with PBD using a hierarchical architecture comprised of graph-convolution and LSTM in which the activity type is continuously leveraged to build activity-informed input for concurrent PBD. In this architecture, both modules receive consecutive frames derived from the data sequence using a sliding window. The HAR module's goal is to recognise the activity being performed and relay that information to the PBD module, which can then identify whether or not protective behaviour is present. Although this architecture has achieved the state-of-the-art classification of PBD using continuous data, its trainable parameters are quite large compared to the number of samples with a significant difference in length, which may affect the model's generalisability. Despite advances in classifying PBD in the EmoPain database, the proposed approaches only considered one set of labels (\textit{i.e.,} PBD), whereas this database provides multiple sets of labels, including annotation for pain level and pain-related behaviour. As a result, in our study, we not only addressed the shortcomings mentioned above, but we also proposed using the GLOCAL classifier to jointly classify pain level and pain-related behaviour.

%-------------------------------------------------------------------------

\section{Proposed Method}\label{sec:methodology}
\subsection{Multidimensional Time Series}
A time series $\mathcal{T} =\langle \mathbf{s}_{1}, \mathbf{s}_{2}, \cdots, \mathbf{s}_{C} \rangle$ is a time-ordered sequence of vectors. Each vector $\mathbf{s}_{i}$ represents
$k$ features of an entity at a specific time point $t_{i}$, where $1 \leq i \leq C$.

\subsection{Sparsely-connected Autoencoders with GRU}
\begin{figure}[!htbp]
    \centering
    \includegraphics[width=1\linewidth]{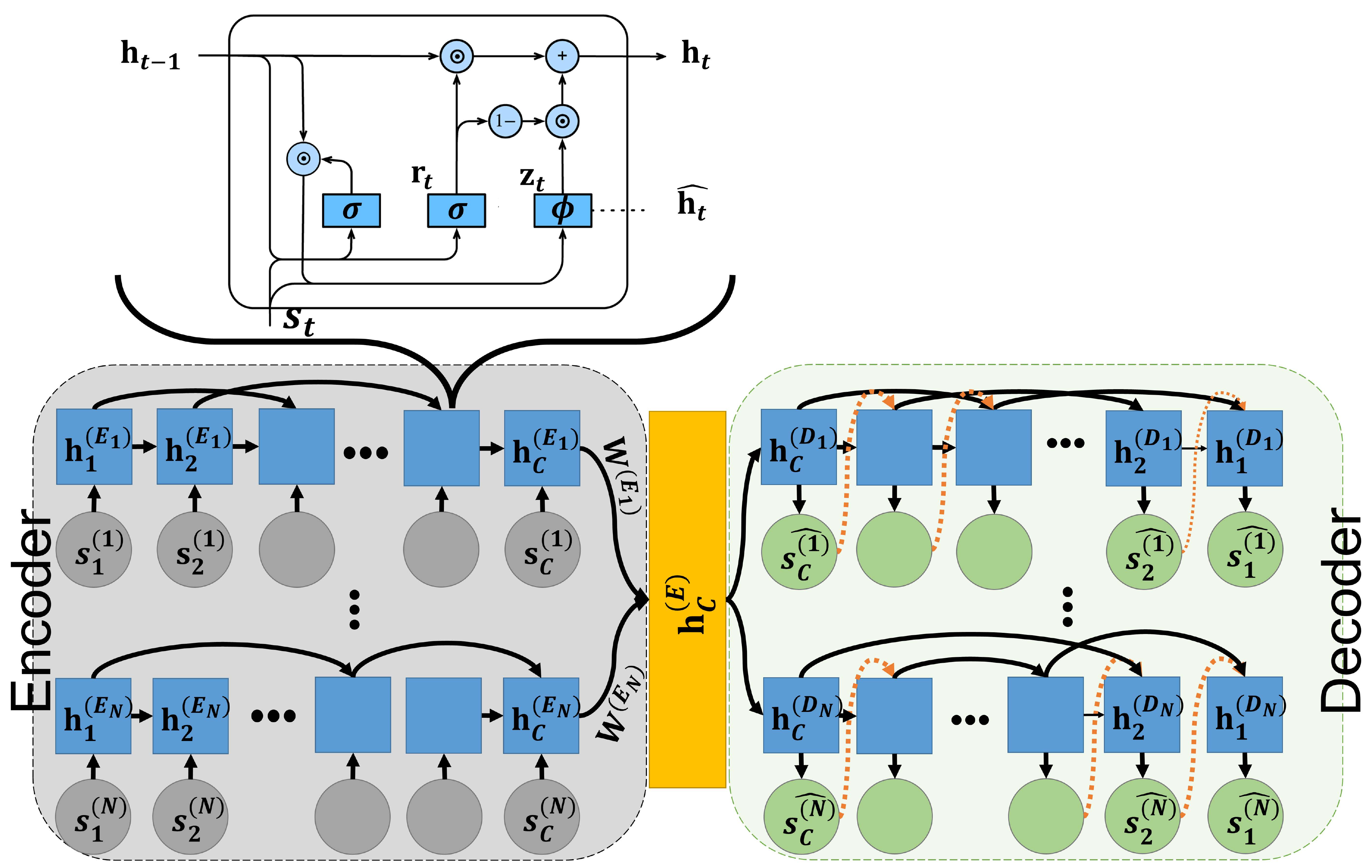}
    \caption{The proposed GRU-based s-RNNs Autoencoder architecture with a shared training framework. This architecture encodes and decodes $N$ input time series to represent the feature space for all inputs in a shared layer $\mathbf{h}_{C}^{(E)}$}.
    \label{fig:GAE}
\end{figure}

Motivated by~\cite{kieu2019outlier}, we proposed a sparsely-connected RNN ensemble (s-RNNs) with the gated recurrent unit (GRU) that incorporates $N$ autoencoders using a shared training framework to represent the feature space for all inputs in a shared layer $\mathbf{h}_{C}^{(E)}$. When dealing with different inputs, the basic framework trains different autoencoders independently without considering their correlation and interaction. However, taking this correlation into account could enhance the reconstruction of inputs and make the representation of the features in the latent space more discriminative. Therefore, motivated by multi-task learning principles~\cite{long2017learning,cirstea2018correlated}, we proposed a shared training framework that incorporates interactions among different autoencoders (see Fig.~\ref{fig:GAE}).

%---PLACEMENT

Each cell in a standard GRU-based encoder, which is initially fed by $\mathbf{s}_{t} \in \mathcal{T}$, performs computation based on Eq.~\ref{eq:101}.

\begin{align}
    \label{eq:101}
    \mathbf{z}_{t} & = \sigma_{g} \left ( \mathbf{W}_{z}\mathbf{s}_{t} + \mathbf{U}_{z}\mathbf{h}^{(E)}_{t-1} + b_{z} \right ) \nonumber \\
    \mathbf{r}_{t} & = \sigma _{g} \left ( \mathbf{W}_{r}\mathbf{s}_{t} + \mathbf{U}_{r}\mathbf{h}^{(E)}_{t-1} + b_{r} \right ) \nonumber \\
    \mathbf{\hat{h}}^{(E)}_{t} & = \phi_{h} \left ( \mathbf{W}_{h}\mathbf{s}_{t}+\mathbf{U}_{h}(\mathbf{r}_{t} \odot \mathbf{h}^{(E)}_{t-1}) + b_{h} \right ) \nonumber \\
    \mathbf{h}^{(E)}_{t} & = \mathbf{z}_{t} \odot \mathbf{\hat{h}}^{(E)}_{t} + (1-\mathbf{z}_{t}) \odot \mathbf{h}^{(E)}_{t-1}
\end{align}
where $\mathbf{h}_{t}$, $\mathbf{\hat{h}}_{t}$, $\mathbf{z}_{t}$, and $\mathbf{r}_{t}$ represent the activation, candidate activation, update gate, and reset gate at time step $t$, respectively. The output for $t=0$ is $\mathbf{h}_{0} = 0$, and $\odot$ denotes the Hadamard product. Also, $\mathbf{W}$, $\mathbf{U}$ represent weight matrices, and $b$ denotes the bias term. The two activation functions used are a sigmoid function, $\sigma_{g}$, and a hyperbolic tangent function, $\phi_{h}$. When the current unit's hidden state $\mathbf{h}_{t}$ is obtained at time step $t$, it is passed into the next unit at time step $t + 1$.

The time series is reconstructed in reverse order in the decoder (\textit{i.e.,} $\mathcal{\hat{T}} = \langle \mathbf{\hat{s}}_{C}, \cdots, \mathbf{\hat{s}}_{2}, \mathbf{\hat{s}}_{1} \rangle$), with the encoder's last hidden state serving as the decoder's first hidden state. The decoder is fed by the previous hidden state $h^{(D)}_{t-1}$ and the previously reconstructed vector $\mathbf{\hat{s}}_{t-1}$ to reconstruct the current vector and compute the current hidden state using Eq.~\ref{eq:102}.
\begin{align}
    \label{eq:102}
    \mathbf{z}^{'}_{t} & = \sigma_{g} \left ( \mathbf{W}_{z}\mathbb{E}\mathbf{\hat{s}}_{t-1} + \mathbf{U}_{z}\mathbf{h}^{(D)}_{t-1} + b_{z} \right ) \nonumber \\
    \mathbf{r}^{'}_{t} & = \sigma _{g} \left ( \mathbf{W}_{r}\mathbb{E}\mathbf{\hat{s}}_{t-1} + \mathbf{U}_{r}\mathbf{h}^{(D)}_{t-1} + b_{r} \right ) \nonumber \\
    \mathbf{\hat{h}}^{(D)}_{t} & = \phi_{h} \left ( \mathbf{W}_{h}\mathbb{E}\mathbf{\hat{s}}_{t-1} + \mathbf{U}_{h}(\mathbf{r}^{'}_{t} \odot \mathbf{h}^{(D)}_{t-1}) + b_{h} \right ) \nonumber \\
    \mathbf{h}^{(D)}_{t} & = \mathbf{z}^{'}_{t} \odot \mathbf{\hat{h}}^{(D)}_{t} + (1-\mathbf{z}^{'}_{t}) \odot \mathbf{h}^{(D)}_{t-1}
\end{align}
where $\mathbb{E}$, $\mathbf{z}^{'}_{t}$, $\mathbf{r}^{'}_{t}$, and $\mathbf{\hat{h}}^{(D)}_{t}$ are the embedding matrix, update gate, reset gate and candidate activation at time step $t$, respectively.

Our proposed architecture is based on a shared training framework that employs recurrent skip connections~\cite{wang2016recurrent} to preserve sparse connections throughout the training phase. In addition to recurrent skip connections, we used auxiliary connections between RNN units ($L$) to consider additional hidden states in the past through the removal of the connection to the immediate previous hidden state to bring sparsity to the AE architecture. Consider $f(\cdot)$ and $f^{'}(\cdot)$ to be the non-linear functions (\textit{i.e.,} hyperbolic tangent) in Eq.~\ref{eq:101} and Eq.~\ref{eq:102}, respectively that represent the hidden state $\mathbf{h}_{t}$ using $\mathbf{s}_{t}$, $\mathbf{h}_{t-1}$, and $\mathbf{h}_{t-L}$. To remove connections between hidden states, we introduce a sparseness weight vector $\mathbf{w}_{t} = (w_{t}^{(f)}, w_{t}^{(f^{'})})$ to regulate which connections should be removed at each time step $t$, where $w_{t}^{(f)} \in \{0, 1\}$ and $w_{t}^{(f^{'})} \in \{0, 1\}$ in such a way that at least one element in $\mathbf{w}_{t}$ is not equal to 0. The computation is formalised in Eq.~\ref{eq:103}.

\begin{equation}
    \label{eq:103}
    \mathbf{h}_{t}=\frac{f(\mathbf{s}_{t},\mathbf{h}_{t-1}) \cdot w_{t}^{(f)} + f^{'}(\mathbf{s}_{t},\mathbf{h}_{t-L}) \cdot w_{t}^{(f^{'})}}{\left \| \mathbf{w}_{t} \right \|_{0}}
\end{equation}
where $\left \| \mathbf{w}_{t} \right \|_{0}$ denotes the number of non-zero elements in vector $\mathbf{w}_{t}$.

As shown in Fig.~\ref{fig:GAE}, different autoencoders interact in the proposed shared training framework to reduce the likelihood of some encoders overfitting to the original time series and to help make decoders robust and less affected by outliers. This interaction occurs at a shared layer ($\mathbf{h}_{C}^{(E)}$) to include all encoders' last hidden states by concatenating these states using linear weight matrices ($\mathbf{W}^{(E_{i})}$) in Eq.~\ref{eq:104}. The shared layer is then used as the initial hidden state in the shared decoder framework to reconstruct the time series.

\begin{equation}
    \label{eq:104}
    \mathbf{h}_{C}^{(E)} = \mathbf{concat}\left ( \mathbf{h}_{C}^{(E_{1})} \cdot \mathbf{W}^{(E_{1})}, \cdots, \mathbf{h}_{C}^{(E_{N})} \cdot \mathbf{W}^{(E_{N})} \right )
\end{equation}

All autoencoders in the shared framework are trained jointly by minimising the objective function $\mathcal{J}$, as given in Eq.~\ref{eq:105}, which calculates the sum of reconstruction errors for each autoencoder and then applies an L1 regularisation term to the shared hidden state.

\begin{align}
    \label{eq:105}
    \mathcal{J} & = \sum_{i=1}^{N} \mathcal{J}_{i} + \lambda \left \| \mathbf{h}_{C}^{(E)} \right \|_{1} \nonumber \\
     & = \sum_{i=1}^{N} \sum_{t=1}^{C} \left \| \mathbf{s}_{t} - \mathbf{\hat{s}}_{t}^{(D_{i})} \right \|^{2}_{2} + \lambda \left \| \mathbf{h}_{C}^{(E)} \right \|_{1}
\end{align}
where $\lambda$ is a weight that regulates the significance of the L1 regularisation, $ \left \| \mathbf{h}_{C}^{(E)} \right \|_{1}$, and the L1 regularisation makes the shared hidden state $\mathbf{h}_{C}^{(E)}$ sparse to minimise overfitting to the original time series and reduce the influence of outliers.

\subsection{Information-theoretic Features}
In several challenging signal processing systems~\cite{dehshibi2021electrical,dehshibi2021stimulating,chiolerio2022olecular} and applications, such as machine learning~\cite{yu2019understanding}, when the data does not follow a Gaussian distribution and the adaptive system is nonlinear, second-order statistics (\textit{e.g.,} variance, correlation, and mean square error) are insufficient to derive adaptive features from the data. Such applications necessitate higher-order statistics of the data, in which the characteristics of linear/nonlinear adaptive signal processing systems, as well as machine learning applications, can be better represented by employing information-theoretic metrics such as entropy, Simpson diversity, expressiveness, and Lempel-Ziv complexity.

The fundamental idea of information theory is that the ``informational value” of data is determined by the degree of uncertainty. If a highly probable event occurs, the data contains very little information; otherwise, the data is much more informative. These higher-order statistics help to lower uncertainty, which is also the goal of machine learning. Therefore, in this study, we calculated information-theoretic complexity measures to characterise spatio-temporal activity patterns in signals to reduce uncertainty and compensate for variations in the temporal dimension that may not be adequately represented in the latent space of s-RNNs.

Because participants carried out the activities in their personalised manner, the acquired signals with equivalent features in $\mathcal{T}$ have varying lengths. To compensate for the variations in the temporal dimension, we used dynamic time warping (DTW)~\cite{sakoe1978dynamic}, in which the sequences are warped non-linearly in the time dimension, and a warping path is generated to align two signals along this path. Consider two time series, $\mathbf{s}$ and $\mathbf{s}^{'}$, of lengths $m$ and $n$, respectively, for which the DTW is calculated using the optimisation problem at the power of $\theta$ in Eq.~\ref{eq:201}.

\begin{equation}
    \label{eq:201}
    DTW_{\theta}(\mathbf{s},\mathbf{s}^{'}) = \min_{\pi \in \mathcal{A}(\mathbf{s},\mathbf{s}^{'})}\left ( \sum_{u,v \in \pi} d(\mathbf{s}_{u},\mathbf{s}^{'}_{v}) \right )^{\frac{1}{\theta}}
\end{equation}
where $\pi$ is a $\kappa$-length alignment path made up of a sequence of $\kappa$ index pairs $\left( (u_{1}, v_{1}),\cdots,(u_{\kappa},v_{\kappa}) \right)$, $\mathcal{A}(\mathbf{s},\mathbf{s}^{'})$ is the set of all admissible paths, and $d(\cdot,\cdot)$ calculates the distance between the $u^{\text{th}}$ sample of $\mathbf{s}$ and the $v^{\text{th}}$ sample of $\mathbf{s}^{'}$. The DTW similarity measure analyses a set of sequences using two time series as input. In our case, because each $\mathbf{s}_{i}$ consists of several time series, one of the inputs must be a representative of $\mathbf{s}_{i}$ that preserves the magnitude of the extremes and the timing features. Therefore, for each $\mathbf{s}_{i}$, we used DTW Barycenter Averaging~\cite{petitjean2011global} to define a reference time series and then calculated DTW.

We extracted an 8-dimensional feature vector from DTW of each time series ($\Delta=\{\delta_{1},\delta_{2}, \cdots, \delta_{8}\}$) and combined it with represented features in the shared hidden state ($\mathbf{h}_{C}^{(E)}$) in a vanilla way before feeding it into the GLOCAL classifier. Shannon entropy~\cite{shannon1948mathematical}, the logarithm of true diversity, was the first attempt to quantify the degree of uncertainty. R\'{e}nyi entropy~\cite{renyi1961measures} expanded on this concept by calculating the logarithm of true diversity based on any value and is used as a diversity index in higher-order statistics. Simpson diversity~\cite{simpson1949measurement} is another measure of diversity that assesses the degree of concentration when individuals are classified. This metric is determined by two key factors, including (1) richness (\textit{i.e.,} the number of different categories of data present in a database) and (2) evenness (\textit{i.e.,} the similarity of the population size of each of the data present). The expressiveness~\cite{alur1990real} is assessed based on the set of all quantifiable activities of interest. The more that can be said absolutely, the greater expressiveness is. We described how hand-crafted features are extracted and explained them mathematically and theoretically as follows.

\begin{enumerate}
    \item \textbf{Shannon entropy} ($\delta_{1}$): Given a discrete random variable $\pi$ with possible outcomes $\{\pi_{1}, \pi_{2}, \cdots, \pi_{\kappa}\}$, which occur with probability $\{p(\pi_{1}), p(\pi_{2}), \cdots, p(\pi_{\kappa})\}$, the Shannon entropy is formally defined as in Eq.~\ref{eq:202}:
    \begin{equation}
        \label{eq:202}
        \delta_{1} = - \sum_{j=1}^{\kappa} p(\pi_{j}) \log \left ( p(\pi_{j}) \right)
    \end{equation}
    \item \textbf{R\'{e}nyi entropy} ($\delta_{2}$): This entropy (see Eq.~\ref{eq:203}) forms the basis of the concept of generalised dimensions, which is important in statistics as an index of diversity. In our experiments, we set $q = 2$.
    \begin{equation}
        \label{eq:203}
        \delta_{2} = \frac{1}{1-q} \left ( \ln \left ( \sum_{j=1}^{\kappa} p(\pi_{j})^{q} \right ) \right)
    \end{equation}
    \item \textbf{Simpson diversity} ($\delta_{3}$): It is calculated as $\delta_{3} = \sum_{j=1}^{\kappa} p(\pi_{j})^{2}$. It measures the degree of concentration when individuals are classified into types. The value of $\delta_{3}$ ranges between 0 and 1, where 1 represents infinite diversity and 0 no diversity.
    \item \textbf{Space filling} ($\delta_{4}$): It is the ratio of non-zero entries in $\pi$ to the total length of signal.
    \item \textbf{Expressiveness} ($\delta_{5}$): It is calculated as the Shannon entropy $\delta_{1}$ divided by Space filling ratio $\delta_{4}$, where it reflects the `economy of diversity'.
    \item \textbf{Lempel--Ziv complexity} ($\delta_{6}$): It is used to assess temporal signal diversity, \textit{i.e.,} compressibility. We used the Kolmogorov complexity algorithm~\cite{kaspar1987easily} to measure the Lempel--Ziv complexity. It is particularly useful as a scalar metric to estimate the bandwidth of random processes and the harmonic variability in quasi-periodic signals.
    \item \textbf{Perturbation complexity index} ($\delta_{7}$): It is defined as the normalised Lempel–-Ziv complexity of the spatio-temporal pattern of a signal by its Shannon entropy.
    \item \textbf{Diversity index} ($\delta_{8}$): It is a quantitative measure that reflects how many different activities of interest are in the acquired signal and that can simultaneously take into account the phylogenetic relations among the activities of interest, such as richness, divergence or evenness.
\end{enumerate}

\subsection{Multi-label Classification}
The EmoPain database is annotated with two sets of labels, amongst others, one for pain level and one for pain-related behaviour. In this study, we employed the multi-label GLOCAL (Multi-Label Learning with Global and Local Label Correlation) classifier~\cite{zhu2018multi}. GLOCAL uses manifold regularisation in order to model global and local label correlations. This learning method uses the label matrix's low-rank decomposition to obtain latent labels. To minimise the reconstruction error in the classifier output, it decomposes the label matrix into two low-rank Laplacian matrices and substitutes missing-label instances with the label correlation. The similarities between labels are explored by enabling similar label predictions to be similar. The adapted global manifold regularisation yields a global label correlation matrix with a positive matrix position if two labels are positively correlated. The same concept is used in local manifold regularisation, but it is applied to $k$ groups identified using the $k$-means algorithm, resulting in $k$ local label correlation matrices. The optimisation problem is formulated by combining global and local manifold regularisation in order to learn global and local label correlations jointly.

Let $\mathbb{C}=\{c_{1}, \cdots, c_{l}\}$ represent the set of $l$ class labels. The $\eta$-dimensional feature vector of an instance is represented by $x \in X = \{\Delta \cup \mathbf{h}_{C}^{(E)}\} \subseteq \mathbb{R}^{\eta}$, and $y \in Y \subseteq \{-1,1\}^{l}$ represents the ground-truth label vector, where $[y]_{j}=1$ if $x$ matches the class label $c_{j}$ and $-1$ otherwise.

%-------------------------------------------------------------------------

\section{Experiments}\label{sec:experiment}

\subsection{EmoPain Database}\label{db}
The EmoPain database~\cite{aung2015automatic} contains data collected from healthy individuals and individuals with CP. The individuals were asked to perform different activity types, including bend-down, one-leg-stand, sit-to-stand, stand-to-sit, and reach-forward, to reflect movements in everyday life activities. It should be noted that individuals participated in two variants of the same activities, \textit{i.e.,} normal and difficult. In the difficult trials, for the sit-to-stand or stand-to-sit, participants were asked to begin the task at the experimenter's instructions; for the one-leg-stand, participants were asked to not just stand on the preferred leg but also the non-preferred one; for the bend-down and reach-forward movements, participants were required to hold 2 and 1 kg dumbbells, respectively, while performing the movements. No instructions on how to perform the movement were provided in either trial, allowing participants to relax or engage in other motions such as stretching, walking, and self-preparation as needed. Indeed, transitions between activities of interest added typical noise to resemble in-the-wild data-gathering settings.

The data was collected using 18 wearable IMU and four sEMG sensors. The IMU sensors recorded Euler angles, which were then converted into position data for 26 full-body anatomical joints using a MATLAB toolbox~\cite{lawrencemocap05}, as described in~\cite{aung2015automatic}. In this paper, we used the angle data (also referred to as \textbf{Joint Angle}) derived from the joint positions and the angular energies (also referred to as \textbf{Joint Energy}) extracted for each of these angles. In addition, we used sEMG sensor-captured muscle activity data from the right and left upper and lower back muscle groups.

The EmoPain database includes self-reports of pain intensity provided after all repetitions of each exercise in the normal or difficult trial were completed. Individuals with CP were the only ones who rated their pain intensity on a range of 0 (\textit{i.e.,} no pain at all) to 10 (\textit{i.e.,} extreme pain). Additionally, the database includes labels for six categories of pain-related behaviour provided continuously over time by four clinicians while they watched video recordings of the movements. They labelled the presence of each behaviour category.

We used data from 17 healthy individuals and 16 individuals with CP who participated in normal activities, as well as data from 23 healthy individuals and 13 individuals with CP who participated in difficult activities. For the pain level label, we used self-reported values from individuals with CP, with this value set to 0 for healthy individuals. For the protective behaviour label, we used 0 to indicate the absence of protective behaviour when clinicians perceived less protective behaviour throughout the trial and 1 to indicate the presence of protective behaviour.

\subsection{Architecture Details}\label{sec:architecture}
Experiments were conducted on a computer with an Intel Core~i9~CPU, 32~GB of RAM, and a GeForce RTX3080 GPU with 12~GB of RAM in the MATLAB~2022a environment using the Deep Learning and Signal Processing toolboxes.

\begin{figure}[!htbp]
    \centering
    \includegraphics[width=1\linewidth]{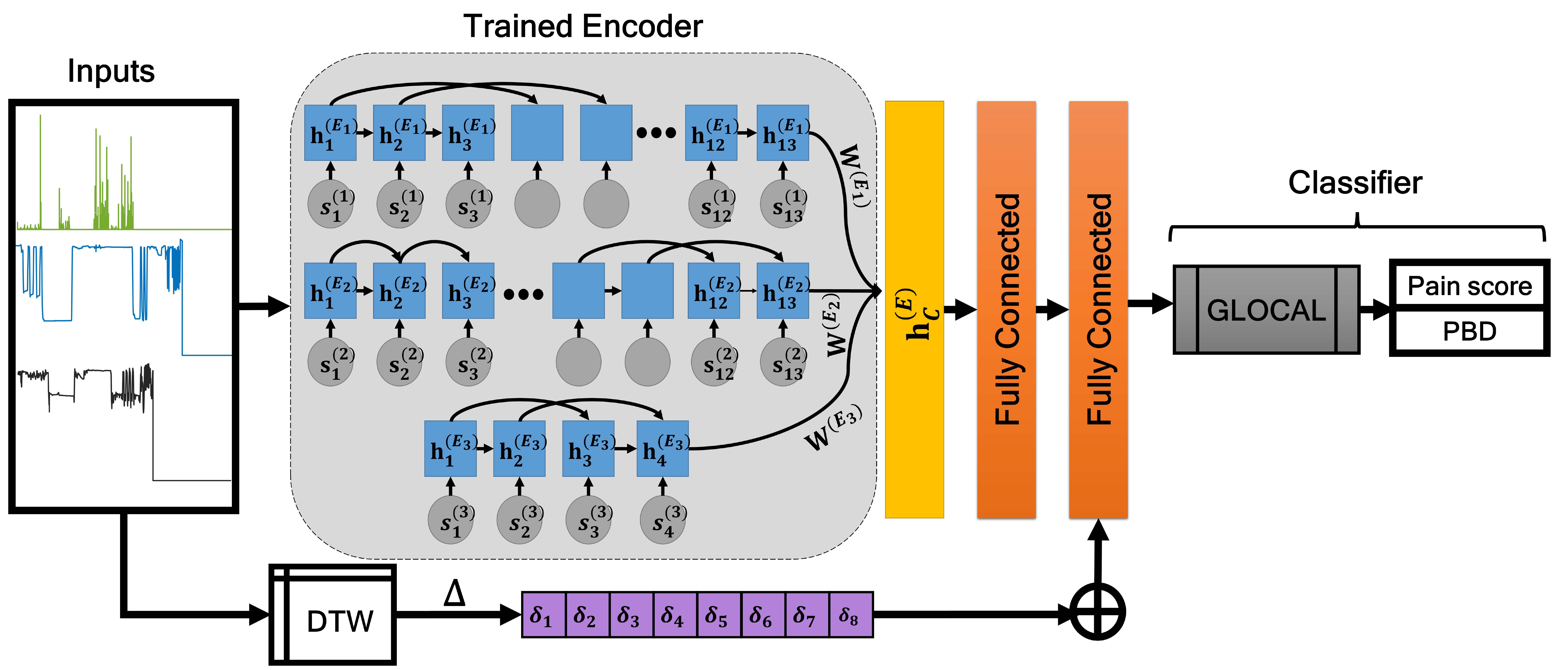}
    \caption{We used the MoCap MATLAB toolbox~\cite{lawrencemocap05} to extract 13 \textbf{Joint Angles} (derived from joint positions) and 13 \textbf{Joint Energies} (angular energies) for each of these Euler angles, which we passed to the encoder as $\mathbf{s}_{1}$ and $\mathbf{s}_{2}$, respectively, while sEMG data sent directly to the encoder as $\mathbf{s}_{3}$. We derive $\mathbf{h}_{C}^{(E)} \in \mathbb{R}^{1 \times 320}$ by feeding data into a trained encoder using a shared training framework. The shared hidden state is then passed to two fully connected layers (FC) with a 0.5 dropout ratio, where the first $\text{FC} \in \mathbb{R}^{1 \times 160}$ and the second $\text{FC} \in \mathbb{R}^{1 \times 80}$. Next, the output of the final fully connected layer is fused with hand-crafted features $\Delta$ using the vanilla fusion approach to represent the body movement of individuals. Finally, this feature vector ($x \in \mathbb{R}^{1 \times 88}$) is fed into the GLOCAL multi-label classifier~\cite{zhu2018multi} to jointly classify pain level and pain-related behaviour. Please note that in this illustration, $\oplus$ represents the concatenation of the high-level descriptor with hand-crafted features, \textbf{Pain score} refers to a value between 0 (no pain) and 10 (severe pain), and \textbf{PBD} refers to no-protective or protective behaviours.}
    \label{fig:architecture}
\end{figure}

The proposed s-RNNs consists of three autoencoders with GRU units. The IMU sensors provide 26 records of $(x,y,z)$ positions in the Euler angle for each participant, while the sEMG sensors provide 4 records of muscle activity. We used the MoCap MATLAB toolbox~\cite{lawrencemocap05} to extract 13 \textbf{Joint Angles} (derived from joint positions) and 13 \textbf{Joint Energies} (angular energies) for each of these Euler angles, which we passed to the encoder as $\mathbf{s}_{1}$ and $\mathbf{s}_{2}$, respectively, while sEMG data passed directly to the encoder as $\mathbf{s}_{3}$. Figure~\ref{fig:architecture} shows the pipeline of the proposed approach to classify pain level and pain-related behaviour.

%---PLACEMENT

%---PLACEMENT
\begin{figure*}[!htbp]
    \centering
    \includegraphics[width=1\linewidth]{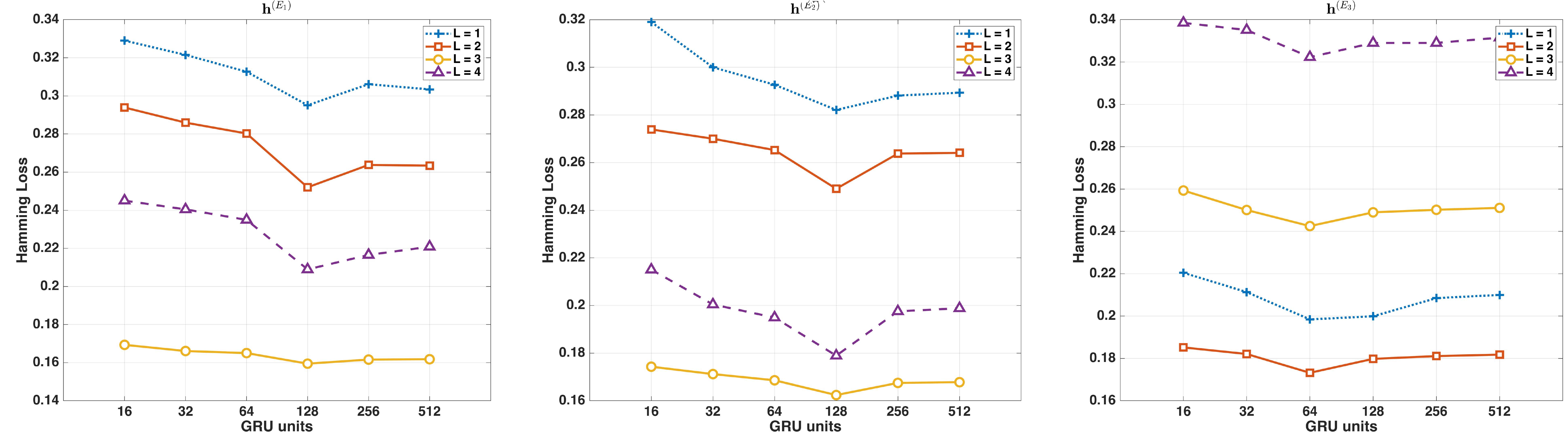}
    \caption{Hamming loss of GLOCAL as a function of (GRU units, $L$) for the encoders. GLOCAL achieved $\text{HL} = 0.159$ with (GRU units,~$L$) = (128, 3) for $\mathbf{h}^{(E_{1})}$, $\text{HL} = 0.162$ with (GRU units,~$L$) = (128, 3) for $\mathbf{h}^{(E_{2})}$, and $\text{HL} = 0.173$ with (GRU units,~$L$) = (64, 2) for $\mathbf{h}^{(E_{3})}$.}
    \label{fig:gru_L_E}
\end{figure*}

The encoder has two hyper-parameters, the number of GRU units and the skip connection jump size $L$, which are later utilised to regulate the pain level and pain-related behaviour classification results using GLOCAL classifier. We trained the proposed architecture as a function of (GRU units, $L$) and plotted the hamming loss (see Section~\ref{sec:metrics}) to identify the best trade-off between these two hyper-parameters for each $\mathbf{h}^{(E_{i})},~ 1 \leq i \leq 3$ (see Fig.~\ref{fig:gru_L_E}). We obtained hamming loss values of $\{0.159, 0.162, 0.173\}$ for $\mathbf{h}^{(E_{1})} \in \mathbb{R}^{128}$, $\mathbf{h}^{(E_{2})}\in \mathbb{R}^{128}$, and $\mathbf{h}^{(E_{3})}\in \mathbb{R}^{64}$, respectively, at $L=\{3,3,2\}$ and used these values in the rest of the experiments. 

We used the hyperbolic tangent function as the non-linear activation function. We evaluated different weight penalty values ($\lambda$) in the range of $[0, 0.005, 0.01, 0.05, 0.1]$ and chose 0.005 for the encoder and decoder. We also chose the sparse weight vector $\mathbf{w}_{t}$ at random. The decoder consists of layers with the same dimensions as the encoder.

We trained the proposed s-RNNs using the Adam optimiser~\cite{kingma2014adam}, which projects the multi-modal data onto a 320-dimensional space ($\eta = 320$). In training the entire network, we used a mini-batch size of 8 that was determined by grid search. The learning rate was set to $10^{-2}$, $10^{-3}$ and $10^{-4}$ for the first 70, next 30, and final 30 epochs, respectively, where the training data was shuffled before each epoch. This policy was used to prevent model divergence caused by unstable gradients. In addition, we set the L2 regularisation, gradient decay factor, and denominator offset to $10^{-4}$, 0.9, and $10^{-8}$, respectively. Finally, we truncated sequences in each mini-batch to the same length as the shortest sequence. This option prevents padding from being introduced at the expense of discarding data, which we compensated for by fusing the hand-crafted features.

We set the learning rates in the trained encoder layers to zero for the classification to use as the feature extractor and avoid overfitting the layers during GLOCAL training on the EmoPain database. We added two fully connected layers with batch normalisation and dropout layers (with a ratio of 0.5) after the shared hidden state to regularise the output feature vector. The hand-crafted features were concatenated with the output of the final fully connected layer before being fed into the GLOCAL multi-label classifier.

\subsection{Evaluation Metrics}\label{sec:metrics}
We employed Hamming Loss, Coverage, Example-Based Accuracy, Ranking Loss, and F-Measure to report the performance of the GLOCAL multi-label classifier, as suggested by Pereira et al.~\cite{pereira2018correlation}. These metrics could prevent the assessment from presenting redundant information.

\begin{itemize}
    \item Hamming Loss (HL) is a normalised metric in which a prediction error (when an incorrect label is predicted) and a missing error (when a relevant label is not predicted) are considered for all classes. It can be calculated by $\text{HL}(H, X) = \frac{1}{N} \sum_{i=1}^{N}\frac{|Y_{i} \otimes \tilde{Y}_{i}|}{|\mathbb{C}|}$, where $H$ is the feature representative model used by the multi-label classifier, $N$ is the number of test data, and $\otimes$ is the symmetrical difference between the two sets, similar to the XOR operation in Boolean logic.
    \item Coverage (Cvg) counts the average steps to be taken in the ranked list of labels to cover all the relevant labels of the example. It can be calculated by $\text{Cvg}(H, X) = \frac{1}{N} \sum_{i=1}^{N} \max(r_{i}(c)) -1$, where $r_{i}(c)$ is the rank position of the label $c$. The most relevant label has the highest rank, and the least relevant label has the lowest rank ($l$).
    \item Example-based Accuracy (EbA) expresses the overall effectiveness of a classifier, given by $\text{EbA}(H, X) = \frac{1}{N} \sum_{i=1}^{N} \frac{|Y_{i} \cap \tilde{Y}_{i}|}{|Y_{i} \cup \tilde{Y}_{i}|}$.
    \item Ranking Loss (Rkl) calculates the frequency of irrelevant labels that are ranked higher than relevant labels, given by $\text{Rkl}(H, X) = \frac{1}{N} \sum_{i=1}^{N} \frac{1}{|Y_{i}||\tilde{Y}_{i}|}|\{(c_{a},c_{b}) : r_{i}(c_{a}) > r_{i}(c_{b}), (c_{a},c_{b}) \in Y_{i} \times \tilde{Y}_{i} \}|$.
    \item F-Measure is the harmonic mean of Precision and Recall, which is calculated by $\text{F1}(H, X) = \frac{1}{N} \sum_{i=1}^{N}\frac{2|Y_{i} \cap \tilde{Y}_{i}|}{|Y_{i}|+|\tilde{Y}_{i}|}$, where we report it in percentage $\text{F1}(\%)$.
\end{itemize}

To better understand classifier performance in detecting pain-related behaviour, we calculated precision, recall, and F1 in the second evaluation setting using Eq.~\ref{eq:301}.

\begin{align}
    \label{eq:301}
    \mathrm{Precision} = \frac{TP}{TP + FP}~, ~~~ \mathrm{Recall} = \frac{TP}{TP + FN}~, \nonumber \\
    F1 = 2\times \frac{\mathrm{Precision} \times \mathrm{Recall}}{\mathrm{Precision} + \mathrm{Recall}}.
\end{align}
where $TP$, $FP$ and $FN$ stand for true positive, false positive and false negative, respectively.

%---PLACEMENT
\begin{table*}[!htbp]
    \centering
    \caption{The performance of the proposed method and baseline approaches for classifying the pain level and pain-related behaviour. We show the mean of metrics at $95\%$ confidence intervals.}
    \label{tbl:01}
    \resizebox{0.8\textwidth}{!}{%
        \begin{tabular}{lccccc}
        \hline
        \textbf{Method}                             & \textbf{HL}              & \textbf{Cvg}             & \textbf{Rkl}             & \textbf{EbA}             & \textbf{F1-Measure (\%)}          \\ \hline
        MiMT~\cite{olugbade2020movement}            & $0.24 \pm 0.04$          & $3.21 \pm 0.18$          & $0.32 \pm 0.08$          & $0.60 \pm 0.10$          & $63.56 \pm 0.87$          \\
        LSTM+GCN~\cite{wang2021leveraging}          & $0.21 \pm 0.03$          & $3.14 \pm 0.23$          & $0.31 \pm 0.06$          & $0.61 \pm 0.12$          & $65.24 \pm 1.20$          \\
        Stacked~LSTM~\cite{wang2019recurrent}       & $0.32 \pm 0.08$          & $3.77 \pm 0.21$          & $0.37 \pm 0.10$          & $0.58 \pm 0.25$          & $59.24 \pm 1.05$          \\
        Dual-stream~LSTM~\cite{wang2021chronic}     & $0.28 \pm 0.05$          & $3.33 \pm 0.14$          & $0.33 \pm 0.06$          & $0.59 \pm 0.08$          & $62.02 \pm 0.55$          \\
        BANet~\cite{wang2019learning}               & $0.36 \pm 0.06$          & $3.90 \pm 0.43$          & $0.40 \pm 0.09$          & $0.56 \pm 0.15$          & $58.36 \pm 0.83$          \\
        Sparse~VAE~\cite{antelmi2019sparse} & $0.39 \pm 0.07$          & $4.13 \pm 0.05$          & $0.42 \pm 0.17$          & $0.54 \pm 0.05$          & $55.80 \pm 0.75$          \\
        Gaussian~VAE~\cite{guo2018multidimensional}         & $0.42 \pm 0.03$          & $4.43 \pm 0.25$          & $0.44 \pm 0.07$          & $0.52 \pm 0.05$          & $53.78 \pm 0.75$          \\
        \textbf{Ours}                           & $\mathbf{0.17 \pm 0.03}$ & $\mathbf{2.28 \pm 0.23}$ & $\mathbf{0.27 \pm 0.05}$ & $\mathbf{0.69 \pm 0.04}$ & $\mathbf{72.21 \pm 0.64}$ \\ \hline
        \end{tabular}%
    }
\end{table*}

\subsection{Experimental Results}
We evaluated the proposed method in two different settings. In the first setting, we tackled the multi-label and multi-class classification task and compared our proposed method to MiMT~\cite{olugbade2020movement}, LSTM+GCN~\cite{wang2021leveraging}, Stacked LSTM~\cite{wang2019recurrent}, Dual-stream LSTM~\cite{wang2021chronic}, BANet~\cite{wang2019learning} that have already been applied to the EmoPain database. In addition to these approaches, we used the Sparse VAE~\cite{guo2018multidimensional} and Gaussian VAE~\cite{antelmi2019sparse} to extract features for this classification task. We used the cross-validation strategy (leave-one-subject-out) to handle the small sample size. To minimise statistical uncertainty, the results were averaged across independent repetitions for all methods and reported with a 95\% confidence interval.

The results for the multi-label classification in the first evaluation setting are shown in Table~\ref{tbl:01}. The proposed method outperformed the state-of-the-art results by Wang et al.~\cite{wang2021leveraging}, which used the combination of LSTM and GCN, achieving an F1-Measure of 65.24\%, and Olugbade et al. (MiMT)~\cite{olugbade2020movement}, which used distributed time encoding of low-level movement features and joint prediction to achieve an F1-Measure of 63.56\% in predicting the pain level and pain-related behaviour simultaneously.

Rather than performing a multi-label classification task in the second evaluation setting, we focused on evaluating pain-related behaviour in the EmoPain database, as in the other baseline approaches. For this reason, we replaced the GLOCAL classifier with a Softmax with two classes (\textit{i.e.,} non-protective, protective) and reported Precision, Recall, and F1 using the leave-one-subject-out cross-validation strategy. Table~\ref{tbl:02} shows the results for this evaluation setting. 

%---PLACEMENT
\begin{table}[!htbp]
    \centering
    \caption{Comparison of the proposed GRU-based s-RNNs with a shared training framework and baseline approaches for predicting pain-related behaviour in the EmoPain database.}
    \label{tbl:02}
    \resizebox{\columnwidth}{!}{%
    \begin{tabular}{lccc}
        \hline
        \textbf{Method}                             & \textbf{Precision (\%)} & \textbf{Recall (\%)} & \textbf{F1 (\%)} \\ \hline
        MiMT~\cite{olugbade2020movement}            & 77.78                   & 82.35                & 80.00            \\
        LSTM+GCN~\cite{wang2021leveraging}          & 78.87                   & \textbf{83.58}       & 81.16            \\
        Stacked~LSTM~\cite{wang2019recurrent}       & 76.39                   & 80.88                & 78.57            \\
        Dual-stream~LSTM~\cite{wang2021chronic}     & 77.46                   & 80.88                & 79.14            \\
        BANet~\cite{wang2019learning}               & 76.06                   & 79.41                & 77.70            \\
        Sparse~VAE~\cite{antelmi2019sparse}         & 71.83                   & 76.12                & 73.91            \\
        Gaussian~VAE~\cite{guo2018multidimensional} & 67.14                   & 72.31                & 69.63            \\
        \textbf{Ours}                           & \textbf{86.11}          & 81.58                & \textbf{83.78}   \\ \hline
    \end{tabular}%
    }
\end{table}

We achieved a 2.62\% performance improvement over the best-performing benchmarking method~\cite{wang2021leveraging} on the EmoPain database. The superior performance of our method is attributable to (1) the usage of a shared training framework in the ensemble of RNNs and (2) concatenating the hand-crafted features extracted using an information-theoretic approach. This architecture allows the model to learn deeper part-whole relationships and selectively focus on the most informative body movement representations, making our model more robust to the challenges associated with movement in individuals with CP.

\subsection{Ablation Study}
In this section, we examine how hand-crafted features affect the performance of multi-label and single-label classification tasks, as well as the difference between independent (IF) and shared (SF) training frameworks (see Tables~\ref{tbl:03} and~\ref{tbl:04}). Details of this experiment are as follows:

%---PLACEMENT
\begin{table}[!htbp]
\centering
\caption{Comparison of the proposed GRU-based s-RNNs architecture for predicting pain-related behaviour. Here, HC, SF and IF are abbreviations for hand-crafted features, shared training framework and independent training framework, respectively. Also, $\oplus$ and $\circleddash$ mean with and without fusing HC.}
\label{tbl:03}
\resizebox{0.9\columnwidth}{!}{%
    \begin{tabular}{lccc}
        \hline
        \textbf{Method}                                   & \textbf{Precision (\%)} & \textbf{Recall (\%)} & \textbf{F1 (\%)} \\ \hline
        \multicolumn{4}{l}{\cellcolor[HTML]{C0C0C0}Ablation study}                                                            \\ \hline
        s-RNNs (SF) $\circleddash$ HC    & 78.95                   & 81.08                & 80.00            \\
        s-RNNs (IF) $\oplus$ HC          & 77.63                   & 79.73                & 78.67            \\
        s-RNNs (IF) $\circleddash$ HC    & 75.34                   & 76.39                & 75.86            \\ \hline
        \multicolumn{4}{l}{\cellcolor[HTML]{C0C0C0}Proposed architecture}                                                     \\ \hline
        \textbf{Ours} & \textbf{86.11}          & \textbf{81.58}                & \textbf{83.78}   \\ \hline
    \end{tabular}%
}
\end{table}

%---PLACEMENT
\begin{table*}[!b]
\centering
\caption{Comparison of the proposed GRU-based sparsely connected RNNs ensemble with a shared training framework to the proposed architecture for predicting pain level and pain-related behaviour using GLOCAL. Here, HC, SF and IF are abbreviations for hand-crafted features, shared training framework and independent training framework, respectively, and we show the mean of metrics at $95\%$ confidence intervals. Also, $\oplus$ and $\circleddash$ mean with and without fusing HC.}
\label{tbl:04}
\resizebox{0.75\textwidth}{!}{%
    \begin{tabular}{lccccc}
        \hline
        \textbf{Method}             & \textbf{HL}             & \textbf{Cvg}             & \textbf{Rkl}             & \textbf{EbA}             & \textbf{F1-Measure (\%)}  \\ \hline
        \multicolumn{6}{l}{\cellcolor[HTML]{C0C0C0}Ablation study}                                                                                                         \\ \hline
        Ours (SF) $\circleddash$ HC & $0.20 \pm 0.08$         & $2.85 \pm 0.05$          & $0.30 \pm 0.10$          & $0.64 \pm 0.20$          & $67.12 \pm 0.83$          \\
        Ours (IF) $\oplus$ HC       & $0.25 \pm 0.03$         & $3.45 \pm 0.12$          & $0.34 \pm 0.08$          & $0.60 \pm 0.15$          & $63.24 \pm 1.05$          \\
        Ours (IF) $\circleddash$ HC & $0.33 \pm 0.12$         & $3.92 \pm 0.38$          & $0.38 \pm 0.15$          & $0.57 \pm 0.63$          & $59.35 \pm 0.9$           \\ \hline
        \multicolumn{6}{l}{\cellcolor[HTML]{C0C0C0}Proposed architecture}                                                                                                  \\
        \textbf{Ours}               & $\mathbf{0.17 \pm 0.03}$ & $\mathbf{2.28 \pm 0.23}$ & $\mathbf{0.27 \pm 0.05}$ & $\mathbf{0.69 \pm 0.04}$ & $\mathbf{72.21 \pm 0.64}$ \\ \hline
    \end{tabular}%
}
\end{table*}
\begin{enumerate}
    \item We excluded the hand-crafted features from the proposed s-RNNs. In this way, GLOCAL and Softmax classifiers were thus solely fed by represented features in $\mathbf{h}^{(E)}_{C} \in \mathbb{R}^{1 \times 320}$. The results in Tables~\ref{tbl:03} and~\ref{tbl:04} show that excluding the hand-crafted features decreases the performance of classifiers. The performance drop is attributable to hand-crafted features' potential to enrich the s-RNNs feature space ($\mathbf{h}^{(E)}_{C}$) by compensating for temporal dimension variations that may not be perfectly represented in the latent space of AEs.
    \item We examined differences between IF and SF, ensuring that both use three GRU-based s-RNNs autoencoders and built IF and SF on top of these autoencoders. Furthermore, we studied the impact of the presence and exclusion of hand-crafted features in IF on the performance of both classification tasks.
\end{enumerate}

%---PLACEMENT
\begin{figure*}[!htbp]
    \centering
    \includegraphics[width=1\linewidth]{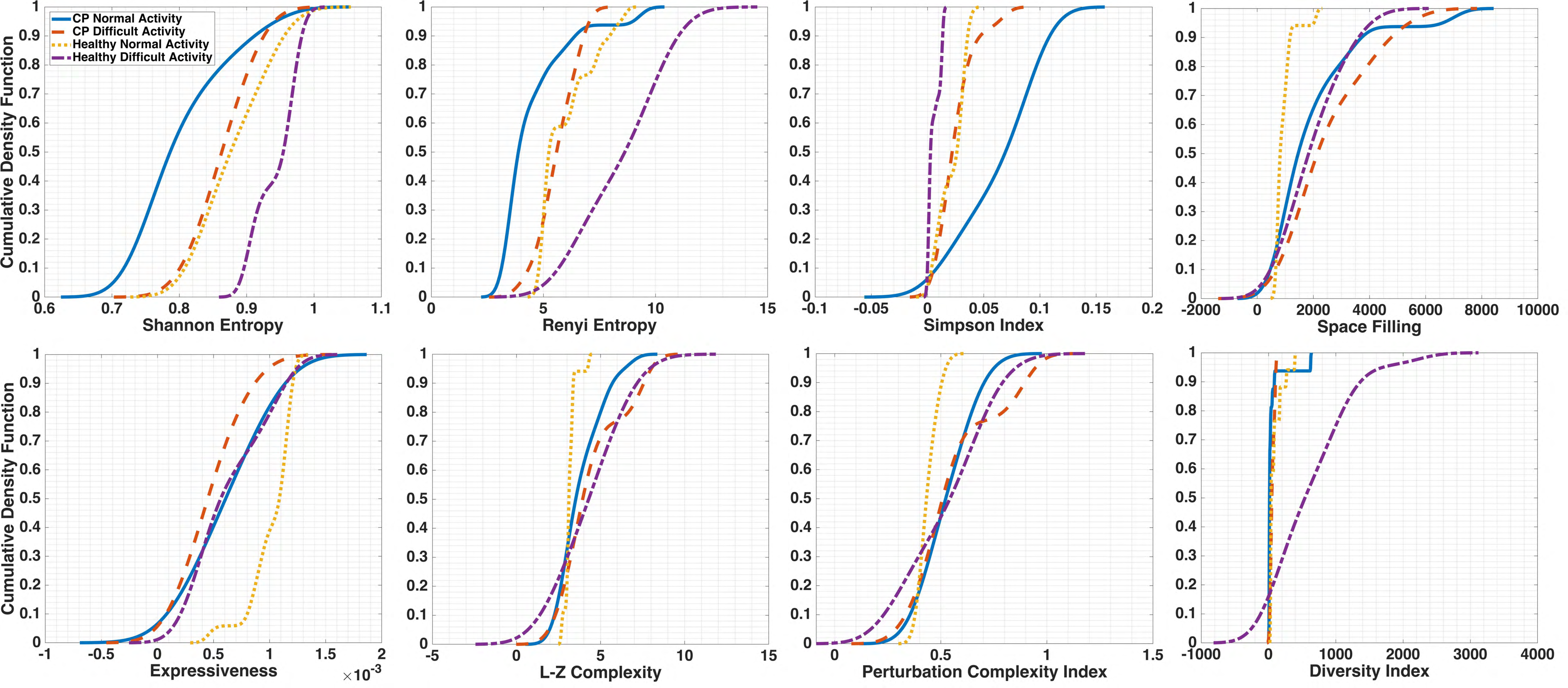}
    \caption{Cumulative distribution function of hand-crafted features.}
    \label{fig:handcrafted}
\end{figure*}

Tables~\ref{tbl:03} and~\ref{tbl:04} indicate that employing the shared training framework and hand-crafted features contributes the most to multi-label and single-label classification tasks. However, excluding hand-crafted features significantly reduces the efficiency of the proposed method for SF and IF. The primary goal of using the proposed s-RNNs is to reduce the dimensionality of input data in a database with a small sample size to deal with the curse of dimensionality and overfitting problems. Furthermore, using the IF training framework mainly without using the hand-crafted features could attenuate the multimodal data semantic relationship in the latent space.

Figure~\ref{fig:handcrafted} shows the cumulative distribution function (CDF) of hand-crafted features to demonstrate their discriminative abilities in normal and difficult activities for both healthy individuals and individuals with CP. We can see that the curves associated with normal and difficult activities are discriminative in all complexity measures derived from information theory.

In particular, the Shannon entropy and the R\'{e}nyi entropy CDFs show similar patterns. The wide range of the Shannon entropy CDF for CP during normal activity may capture two important factors affecting how CP people approach less demanding movements. First of all, many people with CP develop very controlled and specific strategies to deal with perceived physical challenges, which reduces entropy when performing activities. Similarly, disparities in psychological and physical capacity across the group~\cite{wang2019learning} may lead to various strategies used by different people. This is also highlighted by the wide range of Simpson Index CDF values for CP participants engaged in normal activity.

At the same time, participants with CP have a consistently higher level of entropy in difficult activities. As anxiety increases, the way and extent to which people engage in movement (if at all) becomes highly variable across people, as this is dictated less by perceived capability but more by perceived danger. The higher entropy in healthy participants performing the difficult activity is interesting. This higher level of entropy, when compared to the ``CP Difficult Activity" group, can be explained by the fact that all healthy people engage in more demanding everyday activities rather than refraining from doing so, with the caveat of increasing variety in movement trajectory (\textit{e.g.,} amount of bending of the knee and movement speed as they do full bend down to pick up weights from the floor and return or level of balance). This is also supported by the wide range of CDF Index of Diversity values for ``Healthy Difficult Activity".  

The Expressiveness CDF is also particularly interesting. Indeed, only the CDF for ``Healthy Normal Activity" has a higher level of expressiveness (idiosyncrasy), suggesting ease of movement execution for that case. Instead, the CDFs for ``Healthy Difficult Activity" and both CDFs for participants with CP show a broader spectrum of expressiveness, suggesting the greater difficulty of movement execution and reduced idiosyncrasy, possibly for the sake of optimising execution for the healthy participants or addressing anxiety for participants with CP.
%-------------------------------------------------------------------------

\section{Conclusion}\label{sec:conclusion}
Chronic pain is a prevalent long-term condition that interferes with the everyday physical functioning of millions of people worldwide~\cite{rice2016pain,kamerman2020almost,yong2022prevalence}. The body movements of those with CP can influence biometrics analysis. This is because CP can cause changes in muscle tension and posture, which can, in turn, affect the way that algorithms interpret biometric data. These changes can be subtle, such as how people walk, or more pronounced, such as how people hold their heads.

Deep learning-based biometric algorithms have been shown to be more accurate than traditional biometrics methods. Although deep learning algorithms can be fooled by changes in body movement (\textit{e.g.,} protective behaviours of people with chronic pain), they can be trained to account for these changes to improve the accuracy of biometrics analysis. Because there is no dedicated benchmark to investigate the impact of the protective behaviours of individuals suffering from CP on the performance of biometric analysis algorithms, we concentrated primarily on detecting protective behaviours in this study. Indeed, we believe that the results of this study and the findings of previous studies in this area can pave the way for further research into the effect of protective behaviours on the performance of biometric analysis algorithms.

In this paper, we proposed a sparsely-connected recurrent neural networks (s-RNNs) ensemble with the gated recurrent unit (GRU) that incorporates multiple autoencoders using a shared training framework for identifying pain level and pain-related behaviour in the EmoPain database~\cite{aung2015automatic}. The proposed architecture employed recurrent skip connections to maintain sparse connections throughout the training phase, in which auxiliary connections between GRUs were used to consider additional hidden states in the past when the immediate connection to the previous hidden state was removed. Furthermore, different autoencoders interact in the shared training framework to reduce the probability of encoders overfitting to the original time series and help make decoders robust and less impacted by outliers. To compensate for variations in the temporal dimension that may not be perfectly represented in the latent space of s-RNNs, we also propose fusing hand-crafted features extracted from time series using the information-theoretic approaches with represented features in the shared hidden state of s-RNNs.

We evaluated the proposed approach in two separate settings. In the first setting, we tackled the multi-label classification task and compared our proposed method to baseline approaches applied to the EmoPain database. In the second evaluation setting, we focused solely on evaluating pain-related behaviour in the EmoPain database, as in the previous baseline approaches. Our experimental results showed that our proposed method outperformed the state-of-the-art in both multi-label and single-label classification tasks. 

It should be noted that the proposed s-RNNs architecture can be used to analyse a variety of multidimensional input types, such as skeleton tracking from videos (\textit{e.g.,} through OpenPose~\cite{cao2021open} and Kinect~\cite{franco2020multimodal}) or directly 2D/3D computer vision-based body movement tracking data. While the skeleton joints could be used as input to the s-RNNs for the former, using 2D/3D video raw data would require some architectural refinement. For instance, this could be attained by changing the internal modules of the proposed s-RNNs (\textit{e.g.,} substituting GRU with convolution, transformer or diffusion modules) to handle the 2D/3D image or video inputs. Nevertheless, the rationale for using wearable sensors is that chronic pain rehabilitation occurs primarily during functional activity, such as chores at home, work, going out in the city to visit museums, shopping, and playing with kids in the park. As such, a video-based method would be constrained since it requires being in front of a video camera. These recommendations are based on studies conducted with patients and clinicians~\cite{papi2015knee,papi2016wearable,singh2017supporting,olugbade2019can}. Furthermore, video-based tracking poses considerable ethical and privacy concerns, mainly when used outside one's home.

In conclusion, by analysing the body movements of individuals with chronic pain with deep learning, we may better understand the mechanisms underlying this condition. This is something that should be taken into account when using these methods for developing biometrics applications for security or healthcare. This could be used, for instance, to identify areas of a city that are less frequented by people with chronic pain and increase accessibility in such regions. In addition, they could be used to adapt the computation of people's own biometrics in the context of gait modification to prevent falling or transitionally altered behaviour to prevalent and fluctuating conditions such as chronic musculoskeletal pain.
%-------------------------------------------------------------------------

\section*{Acknowledgement}
This project has received funding from the European Research Council (ERC) under the European Union's Horizon 2020 research and innovation programme, with grant agreement No. 101002711.

\bibliographystyle{IEEEtran}
\typeout{}
\bibliography{reference.bib}

\begin{thebibliography}{10}\itemsep=-1pt

\bibitem{adhane2021deep}
Gereziher Adhane, Mohammad~Mahdi Dehshibi, and David Masip.
\newblock {A Deep Convolutional Neural Network for Classification of Aedes
  Albopictus Mosquitoes}.
\newblock {\em IEEE Access}, 9:72681--72690, 2021.

\bibitem{adhane2021use}
Gereziher Adhane, Mohammad~Mahdi Dehshibi, and David Masip.
\newblock {On the Use of Uncertainty in Classifying Aedes Albopictus
  Mosquitoes}.
\newblock {\em IEEE Journal of Selected Topics in Signal Processing},
  16(2):224--233, 2022.

\bibitem{alur1990real}
Rajeev Alur and Thomas~A. Henzinger.
\newblock Real-time logics: complexity and expressiveness.
\newblock In {\em {[1990] Proceedings. Fifth Annual IEEE Symposium on Logic in
  Computer Science}}, pages 390--401. IEEE, 1990.

\bibitem{antelmi2019sparse}
Luigi Antelmi, Nicholas Ayache, Philippe Robert, and Marco Lorenzi.
\newblock {Sparse Multi-Channel Variational Autoencoder for the Joint Analysis
  of Heterogeneous Data}.
\newblock In {\em {Proceedings of the 36th International Conference on Machine
  Learning}}, pages 302--311. PMLR, 2019.

\bibitem{ashtari2022amulti}
Mona Ashtari-Majlan, Abbas Seifi, and Mohammad~Mahdi Dehshibi.
\newblock {A Multi-Stream Convolutional Neural Network for Classification of
  Progressive MCI in Alzheimer's Disease Using Structural MRI Images}.
\newblock {\em {IEEE Journal of Biomedical and Health Informatics}},
  26(8):3918--3926, 2022.

\bibitem{aung2015automatic}
Min S.~H. Aung, Sebastian Kaltwang, Bernardino Romera-Paredes, Brais Martinez,
  Aneesha Singh, Matteo Cella, Michel Valstar, Hongying Meng, Andrew Kemp,
  Moshen Shafizadeh, Aaron~C. Elkins, Natalie Kanakam, Amschel de Rothschild,
  Nick Tyler, Paul~J. Watson, Amanda C. de~C. Williams, Maja Pantic, and Nadia
  Bianchi-Berthouze.
\newblock {The Automatic Detection of Chronic Pain-Related Expression:
  Requirements, Challenges and the Multimodal EmoPain Dataset}.
\newblock {\em {IEEE Transactions on Affective Computing}}, 7(4):435--451,
  2016.

\bibitem{bhanu2017deep}
Bir Bhanu, Ajay Kumar, et~al.
\newblock {\em {Deep Learning for Biometrics}}.
\newblock Springer Cham, 2017.

\bibitem{cao2021open}
Zhe Cao, Gines Hidalgo, Tomas Simon, Shih-En Wei, and Yaser Sheikh.
\newblock {OpenPose: Realtime Multi-Person 2D Pose Estimation Using Part
  Affinity Fields}.
\newblock {\em {IEEE Transactions on Pattern Analysis and Machine
  Intelligence}}, 43(11):172--186, 2021.

\bibitem{chiolerio2022olecular}
Alessandro Chiolerio, Mohammad~Mahdi Dehshibi, Giuseppe Vitiello, and Andrew
  Adamatzky.
\newblock {Molecular Collective Response and Dynamical Symmetry Properties in
  Biopotentials of Superior Plants: Experimental Observations and Quantum Field
  Theory Modeling}.
\newblock {\em Symmetry}, 14(9):1792:1--19, 2022.

\bibitem{cirstea2018correlated}
Razvan-Gabriel Cirstea, Darius-Valer Micu, Gabriel-Marcel Muresan, Chenjuan
  Guo, and Bin Yang.
\newblock {Correlated Time Series Forecasting Using Multi-Task Deep Neural
  Networks}.
\newblock In {\em {Proceedings of the 27th ACM International Conference on
  Information and Knowledge Management}}, pages 1527--1530. Association for
  Computing Machinery, 2018.

\bibitem{collins2015reducing}
Steven~H. Collins, M.~Bruce Wiggin, and Gregory~S. Sawicki.
\newblock Reducing the energy cost of human walking using an unpowered
  exoskeleton.
\newblock {\em Nature}, 522(7555):212--215, 2015.

\bibitem{dehshibi2021electrical}
Mohammad~Mahdi Dehshibi and Andrew Adamatzky.
\newblock {Electrical activity of fungi: Spikes detection and complexity
  analysis}.
\newblock {\em Biosystems}, 203:104373, 2021.

\bibitem{dehshibi2021deep}
Mohammad~Mahdi Dehshibi, Bita Baiani, Gerard Pons, and David Masip.
\newblock {A deep multimodal learning approach to perceive basic needs of
  humans from Instagram profile}.
\newblock {\em IEEE Transactions on Affective Computing}, pages 1--14, 2021.

\bibitem{dehshibi2021stimulating}
Mohammad~Mahdi Dehshibi, Alessandro Chiolerio, Anna Nikolaidou, Richard Mayne,
  Antoni Gandia, Mona Ashtari-Majlan, and Andrew Adamatzky.
\newblock {Stimulating Fungi Pleurotus ostreatus with Hydrocortisone}.
\newblock {\em {ACS Biomaterials Science \& Engineering}}, 7(8):3718--3726,
  2021.

\bibitem{dehshibi2019cubic}
Mohammad~Mahdi Dehshibi and Jamshid Shanbehzadeh.
\newblock {Cubic norm and kernel-based bi-directional PCA: toward age-aware
  facial kinship verification}.
\newblock {\em The Visual Computer}, 35:23--40, 2019.

\bibitem{franco2020multimodal}
Annalisa Franco, Antonio Magnani, and Dario Maio.
\newblock A multimodal approach for human activity recognition based on
  skeleton and rgb data.
\newblock {\em {Pattern Recognition Letters}}, 131:293--299, 2020.

\bibitem{guo2020gluoncv}
Jian Guo, He He, Tong He, Leonard Lausen, Mu Li, Haibin Lin, Xingjian Shi,
  Chenguang Wang, Junyuan Xie, Sheng Zha, Aston Zhang, Hang Zhang, Zhi Zhang,
  Zhongyue Zhang, Shuai Zheng, and Yi Zhu.
\newblock {GluonCV and GluonNLP: Deep Learning in Computer Vision and Natural
  Language Processing}.
\newblock {\em {Journal of Machine Learning Research}}, 21(23):1--7, 2020.

\bibitem{guo2018multidimensional}
Yifan Guo, Weixian Liao, Qianlong Wang, Lixing Yu, Tianxi Ji, and Pan Li.
\newblock {Multidimensional Time Series Anomaly Detection: A GRU-based Gaussian
  Mixture Variational Autoencoder Approach}.
\newblock In {\em {Proceedings of The 10th Asian Conference on Machine
  Learning}}, pages 97--112. PMLR, 2018.

\bibitem{hayden2005exercise}
Jill Hayden, Maurits~W. Van~Tulder, Antti Malmivaara, and Bart~W. Koes.
\newblock {Exercise therapy for treatment of non‐specific low back pain}.
\newblock {\em {Cochrane Database of Systematic Reviews}}, 3(3), 2005.

\bibitem{jain201650}
Anil~K. Jain, Karthik Nandakumar, and Arun Ross.
\newblock {50 years of biometric research: Accomplishments, challenges, and
  opportunities}.
\newblock {\em {Pattern Recognition Letters}}, 79:80--105, 2016.

\bibitem{kamerman2020almost}
Peter~R. Kamerman, Debbie Bradshaw, Ria Laubscher, Victoria Pillay-van Wyk,
  Glenda~E. Gray, Duncan Mitchell, and Sean Chetty.
\newblock Almost 1 in 5 south african adults have chronic pain: a prevalence
  study conducted in a large nationally representative sample.
\newblock {\em Pain}, 161(7):1629--1635, 2020.

\bibitem{kaspar1987easily}
F. Kaspar and H.~G. Schuster.
\newblock {Easily calculable measure for the complexity of spatiotemporal
  patterns}.
\newblock {\em {Physical Review A}}, 36(2):842--848, 1987.

\bibitem{keefe1982development}
Francis~J. Keefe and Andrew~R. Block.
\newblock {Development of an observation method for assessing pain behavior in
  chronic low back pain patients}.
\newblock {\em {Behavior Therapy}}, 13(4):363--375, 1982.

\bibitem{kieu2019outlier}
Tung Kieu, Bin Yang, Chenjuan Guo, and Christian~S. Jensen.
\newblock {Outlier Detection for Time Series with Recurrent Autoencoder
  Ensembles}.
\newblock In {\em Proceedings of the Twenty-Eighth International Joint
  Conference on Artificial Intelligence, {IJCAI-19}}, pages 2725--2732, 2019.

\bibitem{kingma2014adam}
Jimmy Kingma, Diederik P.and~Ba.
\newblock {Adam: {A} Method for Stochastic Optimization}.
\newblock In {\em {3rd International Conference on Learning Representations,
  {ICLR} 2015}}, pages 1--15, 2015.

\bibitem{koes2006diagnosis}
Bart~W. Koes, M.W. Van~Tulder, and Siep Thomas.
\newblock {Diagnosis and treatment of low back pain}.
\newblock {\em {BMJ}}, 332(7555):1430--1434, 2006.

\bibitem{kumar2019deep}
Upendra Kumar, Esha Tripathi, Surya~Prakash Tripathi, and Kapil~Kumar Gupta.
\newblock {Deep Learning for Healthcare Biometrics}.
\newblock In {\em Design and Implementation of Healthcare Biometric Systems},
  pages 73--108. IGI Global, 2019.

\bibitem{lawrencemocap05}
Neil~D. Lawrence.
\newblock {MOCAP} toolbox for {MATLAB}.
\newblock
  \url{http://inverseprobability.com/publications/lawrence-mocap05.html}, 2005
  (accessed November 14, 2022).

\bibitem{liu2020ntu}
Jun Liu, Amir Shahroudy, Mauricio Perez, Gang Wang, Ling-Yu Duan, and Alex~C.
  Kot.
\newblock {NTU RGB+D 120: A Large-Scale Benchmark for 3D Human Activity
  Understanding}.
\newblock {\em {IEEE Transactions on Pattern Analysis and Machine
  Intelligence}}, 42(10):2684--2701, 2020.

\bibitem{long2017learning}
Mingsheng Long, Zhangjie Cao, Jianmin Wang, and Philip~S Yu.
\newblock {Learning Multiple Tasks with Multilinear Relationship Networks}.
\newblock In {\em {Advances in Neural Information Processing Systems}}, pages
  1--10. Curran Associates, Inc., 2017.

\bibitem{olugbade2020movement}
Temitayo Olugbade, Nicolas Gold, Amanda C de~C Williams, and Nadia
  Bianchi-Berthouze.
\newblock {A Movement in Multiple Time Neural Network for Automatic Detection
  of Pain Behaviour}.
\newblock In {\em Companion Publication of the 2020 International Conference on
  Multimodal Interaction}, pages 442--445. Association for Computing Machinery,
  2020.

\bibitem{olugbade2015pain}
Temitayo~A. Olugbade, Nadia Bianchi-Berthouze, Nicolai Marquardt, and Amanda~C.
  Williams.
\newblock {Pain level recognition using kinematics and muscle activity for
  physical rehabilitation in chronic pain}.
\newblock In {\em 2015 International Conference on Affective Computing and
  Intelligent Interaction (ACII)}, pages 243--249. IEEE, 2015.

\bibitem{olugbade2019can}
Temitayo~A. Olugbade, Aneesha Singh, Nadia Bianchi-Berthouze, Nicolai
  Marquardt, Min S.~H. Aung, and Amanda C. de~C. Williams.
\newblock {How Can Affect Be Detected and Represented in Technological Support
  for Physical Rehabilitation?}
\newblock {\em ACM Transactions on Computer-Human Interaction}, 26(1):1--29,
  2019.

\bibitem{otter2020survey}
Daniel~W. Otter, Julian~R. Medina, and Jugal~K. Kalita.
\newblock A survey of the usages of deep learning for natural language
  processing.
\newblock {\em IEEE Transactions on Neural Networks and Learning Systems},
  32(2):604--624, 2020.

\bibitem{papi2015knee}
Enrica Papi, Athina Belsi, and Alison~H. McGregor.
\newblock A knee monitoring device and the preferences of patients living with
  osteoarthritis: a qualitative study.
\newblock {\em {BMJ Open}}, 5(9):e007980, 2015.

\bibitem{papi2016wearable}
Enrica Papi, Ged~M. Murtagh, and Alison~H. McGregor.
\newblock Wearable technologies in osteoarthritis: a qualitative study of
  clinicians' preferences.
\newblock {\em {BMJ Open}}, 6(1):e009544, 2016.

\bibitem{pereira2018correlation}
Rafael~B. Pereira, Alexandre Plastino, Bianca Zadrozny, and Luiz~H.C.
  Merschmann.
\newblock {Correlation analysis of performance measures for multi-label
  classification}.
\newblock {\em {Information Processing \& Management}}, 54(3):359--369, 2018.

\bibitem{petitjean2011global}
Fran{\c{c}}ois Petitjean, Alain Ketterlin, and Pierre Gan{\c{c}}arski.
\newblock {A global averaging method for dynamic time warping, with
  applications to clustering}.
\newblock {\em {Pattern Recognition}}, 44(3):678--693, 2011.

\bibitem{renyi1961measures}
Alfr{\'e}d R{\'e}nyi.
\newblock On measures of entropy and information.
\newblock In {\em Proceedings of the 4th Berkeley Symposium on Mathematical
  Statistics and Probability}, pages 547--561, 1961.

\bibitem{rice2016pain}
Andrew~S.C. Rice, Blair~H. Smith, and Fiona~M. Blyth.
\newblock Pain and the global burden of disease.
\newblock {\em Pain}, 157(4):791--796, 2016.

\bibitem{sakoe1978dynamic}
Hiroaki Sakoe and Seibi Chiba.
\newblock {Dynamic programming algorithm optimization for spoken word
  recognition}.
\newblock {\em {IEEE Transactions on Acoustics, Speech, and Signal
  Processing}}, 26(1):43--49, 1978.

\bibitem{shannon1948mathematical}
Claude~Elwood Shannon.
\newblock A mathematical theory of communication.
\newblock {\em {The Bell System Technical Journal}}, 27(3):379--423, 1948.

\bibitem{simpson1949measurement}
Edward~H. Simpson.
\newblock {Measurement of Diversity}.
\newblock {\em Nature}, 163(4148):688--688, 1949.

\bibitem{singh2017supporting}
Aneesha Singh, Nadia Bianchi-Berthouze, and Amanda C. de~C. Williams.
\newblock {Supporting Everyday Function in Chronic Pain Using Wearable
  Technology}.
\newblock In {\em {Proceedings of the 2017 CHI Conference on Human Factors in
  Computing Systems}}, pages 3903--3915. Association for Computing Machinery,
  2017.

\bibitem{sullivan2006influence}
Michael~JL Sullivan, Pascal Thibault, Andr{\'e} Savard, Richard Catchlove, John
  Kozey, and William~D. Stanish.
\newblock {The influence of communication goals and physical demands on
  different dimensions of pain behavior}.
\newblock {\em Pain}, 125(3):270--277, 2006.

\bibitem{sundararajan2018deep}
Kalaivani Sundararajan and Damon~L. Woodard.
\newblock {Deep Learning for Biometrics: A Survey}.
\newblock {\em ACM Comput. Surv.}, 51(3):65: 1--34, 2018.

\bibitem{wallace1999minimum}
Chris~S. Wallace and David~L. Dowe.
\newblock {Minimum Message Length and Kolmogorov Complexity}.
\newblock {\em The Computer Journal}, 42(4):270--283, 1999.

\bibitem{wang2021leveraging}
Chongyang Wang, Yuan Gao, Akhil Mathur, Amanda~C. De~C.~Williams, Nicholas~D.
  Lane, and Nadia Bianchi-Berthouze.
\newblock {Leveraging activity recognition to enable protective behavior
  detection in continuous data}.
\newblock {\em Proceedings of the ACM on Interactive, Mobile, Wearable and
  Ubiquitous Technologies}, 5(2):81:1--27, 2021.

\bibitem{wang2019recurrent}
Chongyang Wang, Temitayo~A. Olugbade, Akhil Mathur, Amanda~C. De~C.~Williams,
  Nicholas~D. Lane, and Nadia Bianchi-Berthouze.
\newblock {Recurrent Network Based Automatic Detection of Chronic Pain
  Protective Behavior Using MoCap and SEMG Data}.
\newblock In {\em Proceedings of the 23rd International Symposium on Wearable
  Computers}, pages 225--230. Association for Computing Machinery, 2019.

\bibitem{wang2021chronic}
Chongyang Wang, Temitayo~A. Olugbade, Akhil Mathur, Amanda C. DE~C. Williams,
  Nicholas~D. Lane, and Nadia Bianchi-Berthouze.
\newblock {Chronic Pain Protective Behavior Detection with Deep Learning}.
\newblock {\em ACM Transactions on Computing for Healthcare}, 2(3):23:1--24,
  2021.

\bibitem{wang2016recurrent}
Yiren Wang and Fei Tian.
\newblock {Recurrent Residual Learning for Sequence Classification}.
\newblock In {\em {Proceedings of the 2016 Conference on Empirical Methods in
  Natural Language Processing}}, pages 938--943. Association for Computational
  Linguistics, 2016.

\bibitem{wang2019learning}
Chongyang Wanga, Min Pengb, Temitayo~A. Olugbadea, Nicholas~D. Lanec, Amanda C.
  de~C. Williamsd, and Nadia Bianchi-Berthouzea.
\newblock {Learning Temporal and Bodily Attention in Protective Movement
  Behavior Detection}.
\newblock In {\em {8th International Conference on Affective Computing and
  Intelligent Interaction Workshops and Demos (ACIIW)}}, pages 324--330. IEEE,
  2019.

\bibitem{yang2012machine}
Mingjing Yang, Huiru Zheng, Haiying Wang, Sally McClean, Jane Hall, and Nigel
  Harris.
\newblock {A machine learning approach to assessing gait patterns for Complex
  Regional Pain Syndrome}.
\newblock {\em Medical Engineering \& Physics}, 34(6):740--746, 2012.

\bibitem{yong2022prevalence}
R.~Jason Yong, Peter~M. Mullins, and Neil Bhattacharyya.
\newblock {Prevalence of chronic pain among adults in the United States}.
\newblock {\em Pain}, 163(2):e328--e332, 2022.

\bibitem{yoo2013interpretation}
Tae~Keun Yoo, Sung~Kean Kim, Soo~Beom Choi, Deog~Young Kim, and Deok~Won Kim.
\newblock {Interpretation of movement during stair ascent for predicting
  severity and prognosis of knee osteoarthritis in elderly women using support
  vector machine}.
\newblock In {\em 2013 35th Annual International Conference of the IEEE
  Engineering in Medicine and Biology Society (EMBC)}, pages 192--196. IEEE,
  2013.

\bibitem{yu2019understanding}
Shujian Yu and Jos\'{e}~C. Pr\'{i}ncipe.
\newblock {Understanding autoencoders with information theoretic concepts}.
\newblock {\em Neural Networks}, 117:104--123, 2019.

\bibitem{zenil2019causal}
Hector Zenil, Narsis~A. Kiani, Allan~A. Zea, and Jesper Tegn{\'e}r.
\newblock {Causal deconvolution by algorithmic generative models}.
\newblock {\em Nature Machine Intelligence}, 1(1):58--66, 2019.

\bibitem{zhu2018multi}
Yue Zhu, James~T. Kwok, and Zhi-Hua Zhou.
\newblock {Multi-Label Learning with Global and Local Label Correlation}.
\newblock {\em IEEE Transactions on Knowledge and Data Engineering},
  30(6):1081--1094, 2018.

\end{thebibliography}

\begin{IEEEbiography}[{\includegraphics[width=1in,height=1.25in,clip,keepaspectratio]{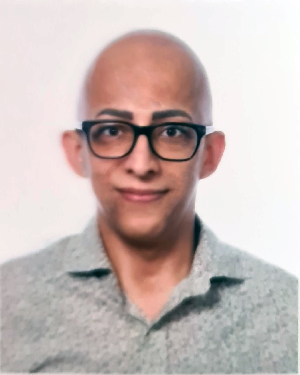}}]{Mohammad Mahdi~Dehshibi} received his PhD in Computer Science in 2017. He is currently a research scientist at Universidad Carlos III de Madrid, Spain. He is also an adjunct researcher at Universitat Oberta de Catalunya (Spain) and the Unconventional Computing Lab. at UWE (Bristol, UK). He has contributed to more than 60 papers published in scientific journals and international conferences. His research interests include Deep Learning, Affective Computing, and Unconventional Computing.
\end{IEEEbiography}

\begin{IEEEbiography}[{\includegraphics[width=1in,height=1.25in,clip,keepaspectratio]{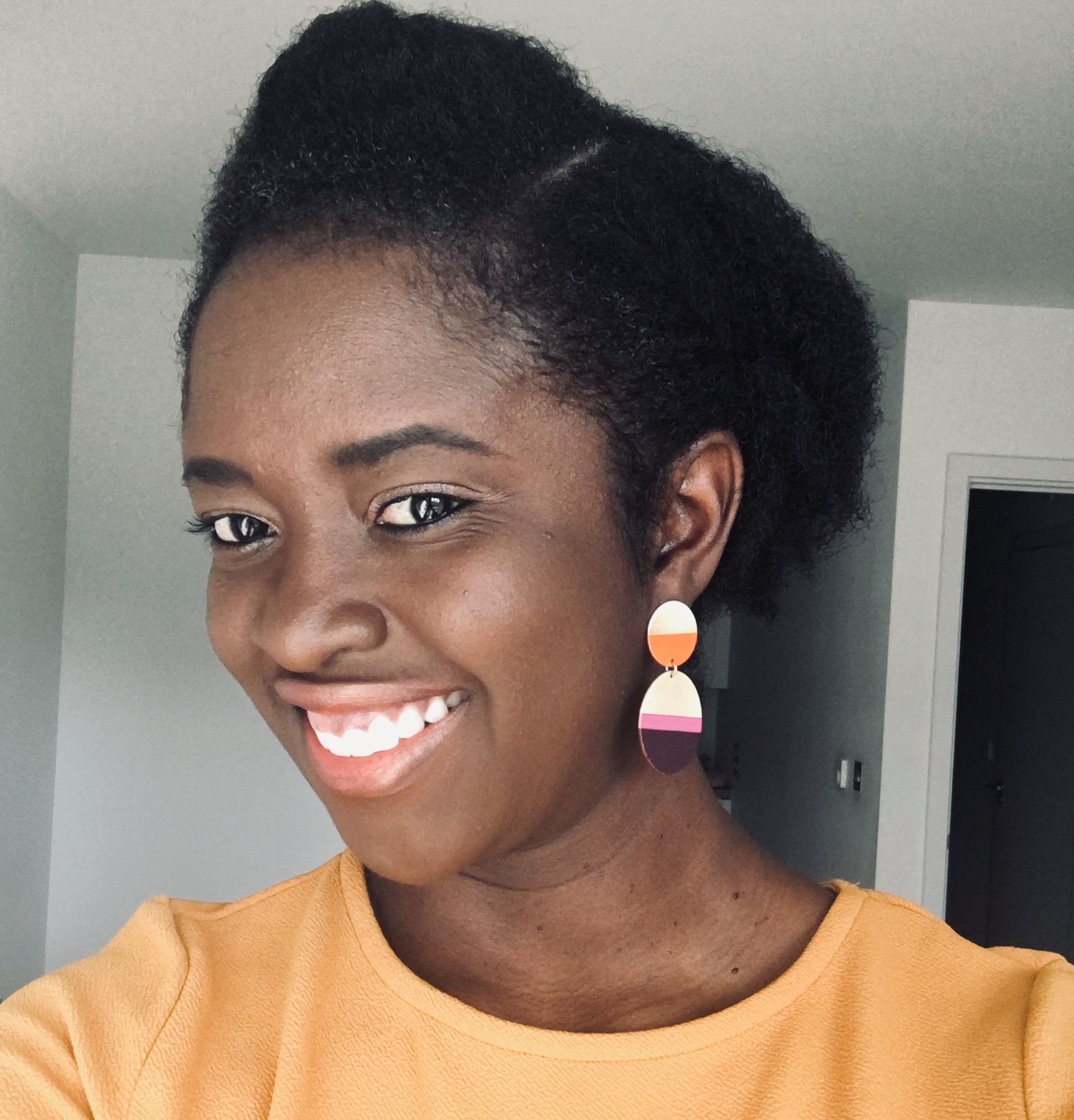}}]{Temitayo~A.~Olugbade} is an applied machine learning researcher at University College London. She has a PhD in Affective Computing, and has strong interests in the (ethical) development of technology that provides support tailored to personal experiences and needs. Beyond core research, she contributes to the development of the IEEE P7014 Standard for Ethical Considerations in Emulated Empathy in Autonomous and Intelligent Systems.
\end{IEEEbiography}

\begin{IEEEbiography}[{\includegraphics[width=1in,height=1.25in,clip,keepaspectratio]{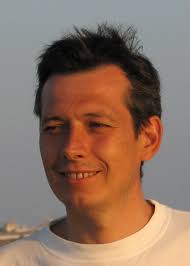}}]{Fernando~Diaz-de-Maria} (Member, IEEE) received the degree in telecommunication engineering and the PhD degree from the Universidad Polit\'{e}cnica de Madrid, Madrid, Spain, in 1991 and 1996, respectively. Since October 1996, he has been an Associate Professor with the Department of Signal Processing and Communications, Universidad Carlos III de Madrid, Madrid. He has co-authored multiple articles in international peer-reviewed journals, several book chapters, and numerous papers at national and international conferences. His current research interests include deep learning, image and video processing, and computer vision. He has led a variety of initiatives and contracts in the mentioned sectors.
\end{IEEEbiography}

\begin{IEEEbiography}[{\includegraphics[width=1in,height=1.25in,clip,keepaspectratio]{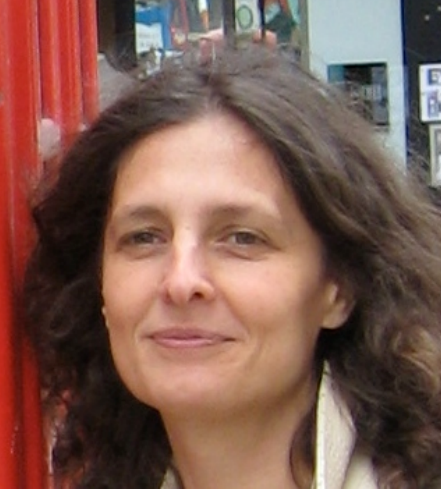}}]{Nadia~Bianchi-Berthouze} is a Professor in Affective Computing and Interaction. She has pioneered the field of Affective Computing from both the machine learning and HCI perspectives. She has published more than 300 papers in affective computing, human-computer interaction and pattern recognition. She has investigated affect-aware technology in  real-life contexts: e.g. EPSRC-funded Emo\&Pain and H2020-funded EnTimeMent on affective-aware rehabilitation technology; EPSRC-funded Technology Circularity Centre on biosensors to capture subjective responses to tactile experiences; and H2020 HU-MAN Manufacturing to measure stress in the industry workfloor contexts.
\end{IEEEbiography}

\begin{IEEEbiography}[{\includegraphics[width=1in,height=1.25in,clip,keepaspectratio]{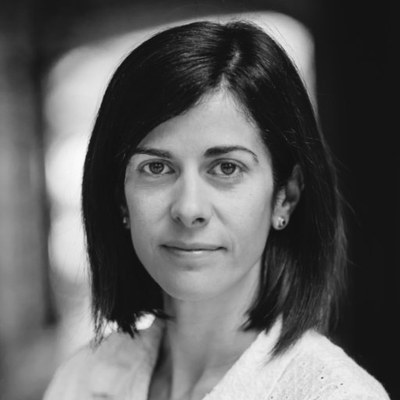}}]{Ana~Tajadura-Jim\'{e}nez} is an Associate Professor at the Department of Computer Science and Engineering of Universidad Carlos III de Madrid (UC3M). She received a degree in telecommunication engineering from the Universidad Polit\'{e}cnica de Madrid, Spain, in 2002 and a PhD degree from Chalmers University of Technology, Spain, in 2008. She has contributed to more than 75 papers in the fields of Psychoacoustics, Cognitive Neuroscience, Affective Computing and Human-Computer Interaction. Her projects, e.g. ESRC-funded The Hearing Body, AEI-funded Magic Shoes and Magic outFit, focus on the use of body tracking and sensory feedback technologies for creating ``Body Transformation Experiences", in which body perception, behaviour, and emotion are modulated, and in their applications for real-life contexts. Her current H2020-funded ERC project  BODYinTRANSIT aims to establish a new fundamental knowledge base about Body Transformation Experiences, using wearable sensor-based sensory feedback devices.
\end{IEEEbiography}
\vfill
\end{document}